\documentclass[aps,prd,twocolumn,10pt,nofootinbib,showkeys,superscriptaddress]{revtex4-1}
\usepackage{amsmath}
\usepackage{amssymb}
\usepackage{graphicx}
\usepackage{color}
\usepackage{hyperref}

\usepackage[english]{babel}
\usepackage[utf8]{inputenc} 

\graphicspath{ {../} }

\RequirePackage{graphicx}
\RequirePackage[english]{babel}
\RequirePackage{amsmath} 
\RequirePackage{verbatim} 
\RequirePackage{mathrsfs}
\RequirePackage{amsfonts}
\RequirePackage{amssymb}
\RequirePackage{hyperref}

\RequirePackage{graphicx}
\RequirePackage{color}

\RequirePackage[utf8]{inputenc} 

\newcommand{\nk}{\textbf{k}}
\newcommand{\nq}{\textbf{q}}
\newcommand{\dphi}{\delta \phi}
\newcommand{\x}{\textbf{x}}
\newcommand{\y}{\textbf{y}}
\newcommand{\bra}{\langle}
\newcommand{\ket}{\rangle}
\newcommand{\mH}{\mathcal{H}}
\newcommand{\mP}{\mathcal{P}}
\newcommand{\mR}{\mathcal{R}}

\newcommand{\nn}{\nonumber \\}

\newcommand{\RI}{\text{R,I}}

\newcommand{\teta}{\tilde{\eta}}
\newcommand{\Hi}{H_{\text{inf}}}

\begin{document}

\title{Enlightening the CSL model landscape in inflation}

\author{Gabriel Le\'{o}n}
\email{gleon@fcaglp.unlp.edu.ar }
\affiliation{Grupo de Cosmolog\'{\i}a, Facultad
	de Ciencias Astron\'{o}micas y Geof\'{\i}sicas, Universidad Nacional de La
	Plata, Paseo del Bosque S/N 1900 La Plata, Argentina.\\
	CONICET, Godoy Cruz 2290, 1425 Ciudad Aut\'onoma de Buenos Aires, Argentina. }

\author{Gabriel R. Bengochea}
\email{gabriel@iafe.uba.ar} \affiliation{Instituto de Astronom\'\i
	a y F\'\i sica del Espacio (IAFE), CONICET - Universidad de Buenos Aires, (1428) Buenos Aires, Argentina}

\begin{abstract}
We propose a novel realization for the natural extrapolation of the continuous spontaneous localization (CSL) model, in order to account for the origin of primordial inhomogeneities during inflation. This particular model is based on three main elements: (i) the semiclassical gravity framework, (ii) a collapse-generating operator associated to a relativistic invariant scalar of the energy-momentum tensor, and (iii) an extension of the CSL parameter(s) as a function of the spacetime curvature. Furthermore, employing standard cosmological perturbation theory at linear order, and for a reasonable range within the parameter space of the model, we obtain a nearly scale invariant power spectrum consistent with recent observational CMB data. This opens a vast landscape of different options for the application of the CSL model to the cosmological context, and possibly sheds light on searches for a full covariant version of the CSL theory.

\end{abstract}
\keywords{Cosmology, Inflation, Quantum Cosmology}

\maketitle

\section{Introduction}

One of the most interesting challenges when trying to combine General Relativity with Quantum theory to explain the very early epoch of the universe, has to do with giving a satisfactory explanation of the quantum-to-classical transition of perturbations originated during inflation. The proposal that, at the very beginning, the universe went through an accelerated phase, called inflation, is extremely successful since it allowed explaining not only some well-known problems of the hot Big Bang model (e.g. the horizon and flatness problems), but it has also offered a way to address the quantum origin of large-scale structures that we see today in the sky, from primordial cosmic seeds \cite{Starobinsky79,Starobinsky80,Guth81,Mukhanov81,Linde82,Linde83,Albrecht82,Bardeen83,Brandenberger84,Hawking82}. The inflationary paradigm predictions for primordial power spectra are strikingly consistent with the most recent accurate observations of cosmic microwave background (CMB) radiation \cite{Planck18a,Planck18b,Planck18c}.

Notice that, the aforementioned quantum to classical transition is closely related to the so-called \emph{measurement problem} in Quantum physics \cite{Wigner63,Omnes,Albert,Maudlin95,Becker,Norsen,Durr,okon14}\footnote{See \cite{Maudlin95}, where the author states the measurement problem
	in a formal and elegant way, and he also describes what are the possible alternatives to address it.}. Let us mention why this problem is notoriously enhanced in the cosmological case \cite{Bell81,PSS06,Sudarsky11,Susana13}\footnote{Non-specialist readers interested in a introductory review on this subject can find it in \cite{Bengochea20}.}. The central point here is that, according to standard Quantum theory, the evolution of any quantum state is always unitary, dictated by the Schr\"odinger equation, which does not break any initial symmetry of the system or destroy quantum superpositions. If at the beginning of inflation the spacetime is assumed to be spatially isotropic and homogeneous, and the perturbations of matter fields (i.e. the inflaton field) were in a quantum vacuum state also invariant under spatial rotations and translations (i.e. the so-called Bunch-Davies vacuum); then one question arises: how do we arrive at an anisotropic and inhomogeneous state (\emph{with} cosmic seeds of structures) from a quantum vacuum state that is perfectly isotropic and homogeneous in space? The traditional early universe paradigm seems to be incomplete in that sense\footnote{The standard explanation for such an emergence is based on the study of the role of \emph{quantum fluctuations} of the scalar field driving the accelerated expansion during the inflationary epoch. However, that analysis lacks a complete description for the problem at hand that is being discussed here \cite{Sudarsky11,Berjon21}.}. And any model that claims to provide a mechanism for the emergence of the seeds of structure must be able to give a convincing answer to the mentioned scenario, without appealing, of course, to the existence of observers or measuring devices which cannot constitute fundamental elements in the very early universe \cite{Hartle93}. There are various proposals that try to explain the quantum-to-classical transition from different approaches. For instance, by invoking the decoherence framework \cite{kiefer09,halliwell,kiefer2,polarski}, it could provide a partial understanding of the issue. However, it was shown that decoherence does not fully addresses the problem \cite{okon16,Adler01,schlosshauer,Sudarsky11,kent,stapp}. Other authors, within the framework of the de Broglie-Bohm theory \cite{bohm}, have also explored applications to this cosmological puzzle, e.g. \cite{Valentini08a,Valentini08b,Neto12,goldstein15,Neto18,Valentini19}.

Known as \emph{objective collapse theories}, they constitute another approach that seeks to address the aforementioned issue, since they were specifically designed to solve the measurement problem in Quantum Mechanics \cite{Pearle76,Ghirardi86,Pearle89,Diosi87,Diosi89,Penrose96}. Reviews on these sort of theories can be found, for instance, in \cite{Bassi1,Bassi2}. The idea of invoking a self-induced collapse in cosmology in order to generate the primordial perturbations and/or accomplish the quantum-to-classical transition has been explored in great detail since 2006 \cite{PSS06}, and it has led to numerous investigations in recent years with varied proposals and results, e.g. \cite{Daniel10,Sudarsky11,Tejedor12,Tejedor12B,Martin12,Pedro13,Das13,Bengochea15,Leon15,Syksy15,Leon16,Stephon16,Leon17,Landau12,Susana13,Benetti16,Bengo17,Ellis18,Pedro18,Benito18,Picci19,Lucila15,Mariani16,Maj17,ModosB,Bouncing16,Josset17,Leon2020,BengoEmer21}.
$\:$In several of those works, a specific version based on the continuous spontaneous localization (CSL) model \cite{Pearle76,Pearle89} adapted to the situation of interest, was chosen to analyze this subject.

The CSL constitutes a particular proposal, based on a non-linear stochastic modification of the standard Schr\"odinger equation, where spontaneous and random collapses occur, resulting in an objective dynamical reduction of the wave function. For the first time, in Ref. \cite{Martin12} the CSL model was applied to the inflationary universe and in Ref. \cite{Pedro13} such exploration was done within the semiclassical gravity framework. The debate on the particular details involved in implementing the CSL model in the cosmological context is still open, and contains an extensive landscape of possibilities constituting an active research area at the moment \cite{MartinShadow,Bengo20Letter,Bengo20Long,Martin20R,Bassi21,Martin21}. As was already pointed out in \cite{Beneficios}, objective collapse models have the appealing feature of connecting plausible resolutions of other open problems in a single unified picture; not only the measurement problem in Quantum Mechanics and the emergence of the primordial inhomogeneities during inflation, but also the problem of time in canonical quantum theories of gravitation, the black hole information paradox \cite{Modak14,Modak15,Modak16} and other applications to cosmology, for instance to account for the dark energy and the late-time accelerated expansion of the universe \cite{Josset17,Corral20,Nucamendi20}.

Recently, in Ref. \cite{Bengo20Long}, a wide variety of open alternatives were discussed when applying the CSL dynamical collapse model to the inflationary era. That exploration included  the important aspects one must face in order to address such a problem. These are: (i) the two different approaches to deal with quantum field theory and gravitation, (ii) the identification of the collapse-generating operator, and (iii) the general nature and values of the CSL model parameters. In that same work, it was also emphasized that all the choices connected with those issues have the potential to dramatically alter the conclusions one can draw.

In this article, we will explore a particular CSL model in the inflationary context, within the vast theoretical landscape available for the extrapolation of the standard CSL (as constructed to deal with non-relativistic many particle Quantum Mechanics) into the realms of relativistic quantum field theory in curved spacetimes. To accomplish this task, we will follow some options raised in Ref.  \cite{Bengo20Long}. In this way, our model possess three main elements which we mention below.

First, one must choose the setting within which one is going to combine quantum field theory (QFT) with gravitation. While in some previous works,  we have explored the use of the CSL using the standard framework of quantization (in which metric and matter fields are quantized simultaneously). Here, as in other previous works, we will adopt the semiclassical gravity framework where gravity is always classical while the matter fields are treated quantum mechanically. This approach has received some known criticisms \cite{Eppley77,Page81}, but those arguments have been discussed and refuted \cite{Mattingly05,Mattingly06,Kent18,Tilloy16,Carlip08,Albers08,Ford05Review,Ford05ReviewBook}; currently,  the theory continues to be of great interest. Moreover, it is not often emphasized that most of the conclusions against are only valid in those contexts where quantum mechanical evolution does not include any sort of measurement-related or spontaneous collapse of the quantum state. We will assume such a framework to be a valid approximation during the inflationary era, which is well after the full quantum gravity regime has ended. This choice is due to the fact that it appears favored from a theoretical and conceptual point of view, in particular, when one wants to incorporate collapse models \cite{Tejedor12,Pedro18,Benito18,Benito20,Bengo20Long}. In fact, it was shown that, within semiclassical gravity, the CSL model predicts generically a very small amplitude of the tensor power spectrum \cite{Lucila15,Maj17,ModosB}, which is very consistent with the current non-detection of B-modes in the CMB data \cite{Planck18b}.

Second, as mentioned in Ref. \cite{Bengo20Long}, there are a large number of operators that are relativistically invariant and do reduce, in the simple non-relativistic regimes encountered in laboratory situations, to the required ``mass density'' or ``energy density''. Some simple examples we might consider are scalars constructed, for example, using the energy-momentum tensor $T_{ab}$. In the present work, for the first time in this line of research, we will explore the consequences of assuming  a collapse generating operator composed of a contraction of the energy-momentum tensor. Specifically, we will choose the scalar $(T_{ab}T^{ab})^{1/2}$  as the collapse operator in our proposal to extrapolate the CSL model into the cosmological context.

Third, we will make a particular choice regarding the parameters of the CSL model. The main idea is the following.

One of the central features of the CSL model, sometimes referred to as the \emph{amplification mechanism}, is that the collapses must be rare for microscopic systems, in order not to alter their quantum behavior as described by the Schr\"odinger equation. But, on the other hand, their effects must increase when several particles are hold together forming a macroscopic system. The amplification mechanism can be characterized through the collapse rate parameter $\lambda$ of the CSL model.

As was discussed at length in \cite{Bengo20Long}, such a parameter used in the non-relativistic versions of the CSL model, let us call it $\lambda_0$, might not be a fundamental constant after all. The original motivation \cite{Pearle76,Ghirardi86,Pearle89,Bassi1,Pearle1994} for choosing the particular value of $\lambda_0 \approx 10^{-16}$ s$^{-1}$ was based on some phenomena occurring at laboratory scales, e.g. the number of atoms in a pointer apparatus, the matter density of nuclear matter, etc. Therefore, it is not obvious at all that there should be any comparison between the value of $\lambda_0$, informed by the existence of hadrons, nuclei and the atomic densities of solids, with parameters at an era when there were no nucleons, no atoms and certainly no solids.

On the other hand, the second parameter of the non-relativistic CSL model, $r_c$, is a length scale characterizing the level of de-localization associated with the onset of the spontaneous collapse. For instance, for a single particle, the effect of the collapse aspect of the modified dynamics is negligible if the width of its wave function is smaller than  $r_c$. But, if it is larger than $r_c$, the collapse mechanism becomes very relevant. The actual numerical value of  $r_c$ comes from experimental constraints using a non-relativistic version of the CSL model. Additionally, for that particular CSL model, the parameter $r_c$ serves to characterize physical processes at laboratory scales. For example, $r_c$ takes into account notions such as atomic dimensions, mean spreads around the equilibrium positions of the lattice points of a crystal, etc. (see Sec. 6.5 of Ref. \cite{Bassi1}). Thus, there are ample grounds to doubt any simple one-on-one connection between the value of $\lambda_0$ and $r_c$, relevant for one regime (involving laboratory scales), with parameters characterizing the model in a completely different one, such as the inflationary universe.

Previous works \cite{Pedro13,Leon16,Bouncing16,Picci19}, where a particular version of the CSL model was applied to inflation, were able to recover the correct shape and amplitude of the CMB spectrum (in addition to being compatible with laboratory constraints). Those results required  making suitable assumptions regarding the dependence of the model parameters on the Fourier mode wavelengths, which in turn, might be viewed as tied to a more fundamental dependence on curvature. However, in the present work, we will make explicit this dependence on the curvature in the CSL parameters, and make no assumption involving a particular decomposition used in the field's modes (e.g. Fourier modes).

As a matter of fact, as suggested by L. Diosi and R. Penrose in e.g. \cite{Diosi84,Diosi87,Diosi89,Penrose96}, it seems quite natural to think that, at a fundamental level, the spontaneous collapse dynamics might be intimately tied with gravitation. Therefore, accepting such a hypothesis, one reasonable option for exploring the relation between the collapse of the wave function and gravitation, is to  replace the collapse parameter of the ordinary CSL model by a function of some geometrical scalar associated with curvature (see e.g. \cite{Beneficios}). In particular, in a series of works devoted to the examination of the famous black hole information puzzle \cite{Modak14,Modak15}, it was found that, during the black hole evaporation, information is lost as a result of the modified evolution described by the CSL model with the additional hypothesis that the collapse rate $\lambda$ is enhanced by the curvature of the spacetime, through the Ricci scalar $R$.

In this paper, we will adopt the same parameterization of $\lambda$ as the authors of Refs. \cite{Modak14,Modak15}, i.e. we will assume that $\lambda$ is a function of $R$ (a more detailed discussion about this point will be given in Sect. \ref{sec_lambdaR}). This idea was first suggested in \cite{Pearle1994,Pearle1995}, and was also one of the options mentioned in \cite{Bengo20Long}. We acknowledge that there are other possibilities for including curvature scalars in the parameterization of $\lambda$. For instance, one could  consider the Kretschmann  scalar $R_{\alpha \beta \mu \nu}R^{\alpha \beta \mu \nu} $, or the scalar $ W_{\alpha \beta \mu \nu}W^{\alpha \beta \mu \nu}\:$, where $W_{\alpha \beta \mu \nu}$ denotes the Weyl tensor.  In fact, as argued in \cite{Beneficios}, the latter choice seems to have the characteristics that could lead to an association of low entropy with the early state of the universe and a large entropy with its late-time state.  That is, the adoption of an objective collapse model with a curvature-dependent collapse parameter involving the Weyl tensor, might result in dynamical-based explanation of the Weyl curvature hypothesis (see Ref.  \cite{Okon16b} and references within for more details). The point is that including an explicit dependence of some curvature scalar in $\lambda$, appears to be a valid option if one accepts the premise that the objective collapse is enhanced by the gravitational degrees of freedom.

The dependence on curvature could result, of course, in an effective temporal dependency of the model parameters. In fact, this kind of effective time dependence (i.e. a time dependence that appears in a certain regime to be a constant, but is in fact a coupling with some other dynamical variable) is one we have already encountered in cosmology; e.g., the masses in the standard model of particles depend on a vacuum expectation value in the Higgs sector, and that field experienced dramatic changes between the inflationary epoch and the present day. So an effective time dependence of the CSL parameters, in the cosmological context, might very well be considered likely.

With respect to the second parameter of the non-relativistic CSL model $r_c$, which is related to the localization of the wave function corresponding to a system of (non-relativistic massive) particles, in this article we will propose a novel approach to generalize such element when analyzing the inflationary universe. In particular, we will assume that, when taking into account the collapse of the wave function corresponding to the inflaton, the quantum two-point correlation function of the field variables is a straightforward way to extrapolate the notion of ``localization'' as given by the ordinary CSL model (we will be more precise in Sec. \ref{sec_smearing}). Recall that, in the latter, the level of localization is encoded in the quantum uncertainty of the position operator.
	
Finally, let us mention that we will base our treatment on a model of spontaneous collapses that is strictly speaking not covariant. We acknowledge that a truly satisfactory proposal, to deal with the subject considered here, should be based on a fully covariant theory. However in absence of such a theory, we can move forward by making reasonable assumptions that could help in finding a complete and workable theory. Fortunately, recent proposals for special relativistic versions of these type of theories indicate that we should be able to address this shortcoming in near future \cite{Tumulka06,Bedingham11,Pearle2015,Modak16}.

Our goal in this paper can be summarized as follows: based on the three main elements mentioned above, and employing standard cosmological perturbation theory at linear order, we will obtain the primordial scalar power spectrum. Also, we will analyze the characteristics needed for the spectrum to be consistent with both the laboratory constraints and those from the CMB observations.

This manuscript is divided in 6 sections plus two Appendices, where in the latter we have included most of the calculations. We start in section \ref{secCSLoriginal} with a brief review of the standard CSL model by discussing the \emph{mass proportional} CSL model. Next, in section \ref{secCSLposta} we describe our proposed framework, in section \ref{secPk} we find an equivalent scalar power spectrum, and in section \ref{secAnalysis} we analyze the features of the obtained spectrum. Finally, in section \ref{conclusions}, we present our conclusions. Regarding conventions and notation, we use a $(-,+,+,+)$ signature for the spacetime metric and units where $c=1=\hbar$.

\section{Brief review of the CSL model}\label{secCSLoriginal}

In this section, we provide a very brief review of the non-relativistic CSL model, so there is no original work here. Specifically, we will focus on the \textit{mass proportional} CSL model. Those readers familiar with the subject can safely skip to the next section. For a pedagogical review, including the main technical aspects, we suggest Ref. \cite{PearleMisc}, and  for a general overview of collapse models see \cite{Bassi1,Bassi2}.

The CSL model is a generalization of the spontaneous collapse model originally proposed by Ghirardi, Rimini and Weber \cite{Ghirardi86} to a system of identical particles. In the CSL model the collapse of the wave function occurs continuously in time, while in the GRW one the collapse happens discretely. The mass proportional  CSL model  has been probed in experiments, where relativistic effects can be ignored (see e.g. Ref. \cite{Bassiexp21}). It consists of two main equations. The first one is a modified version of the Schr\"odinger equation, whose general solution is
\begin{eqnarray}\label{CSLQM}
	|\psi,t\rangle&=&{\hat {\cal T}}e^{-i \int_{0}^{t}dt' \hat H (t') -\frac{1}{4\lambda_0}  \int_{0}^{t}dt' \int d\x' [w(\x,t')-2\lambda_0 \hat A(\x)]^{2}}  \nn
	&\times& |\psi,0\rangle
\end{eqnarray}
Here, $\hat {\cal T}$ is the time-ordering operator and $w(\x,t)$ characterizes a  stochastic process  of white noise type. The probability for the noise $w(\x,t)$ is given by the second main equation, which represents the Probability Rule:
\begin{equation}\label{CSLprobab}
	P(w)dw \equiv\langle\psi,t|\psi,t\rangle\prod_{t_{i}=0}^{t -dt}\frac{dw(\x,t_{i})}{\sqrt{ 2\pi\lambda_0/dt}}
\end{equation}
The state vector norm evolves dynamically, i.e. it is not equal to 1.  In particular, Eq. \eqref{CSLprobab} implies that state vectors with largest norm are the most probable. The total probability is then $\int P(w)dw =1$, provided that $ \bra \psi, 0 |\psi,0\rangle = 1$.

The parameter $\lambda_0$ is the collapse rate for a reference particle in a spatially superposed state, and usually the reference particle is chosen as a nucleon. On the other hand, the operator $\hat A(\x)$ is called the \textit{collapse generating operator}, and for the mass proportional CSL model it is defined as
\begin{equation}\label{Acolop}
	\hat A(\x) = \sum_i  \frac{m_i}{m_0} \frac{1}{(\pi r_c^2)^{3/4}} \int d^3 y \: e^{-|\x-\y |^2/2 r_c^2} \hat N_i(\y)
\end{equation}
where $\hat N_i(\y) = \hat \chi_i^{\dagger} (\y) \hat \chi_i (\y) $  is the particle number density operator, constructed from (non-relativistic) creation an annihilation operators for a particle of type $i$ at $\y$. In addition, $m_i$ is the mass of the corresponding particle species and $m_0$ is the mass of the reference particle, e.g. a nucleon. The second parameter of the model is the smearing length $r_c$, a length scale characterizing the level of de-localization associated with the onset of the spontaneous collapse. In this way, $\hat A(\x)$ is considered as a \textit{smeared mass density} operator at $\x$.

Equation \eqref{CSLQM} describes the complete evolution of an individual system prepared in a given initial state. However, the presence of the noise $w(\x,t)$ clearly  makes the evolution highly random. The only certainty is that, by construction, the system will be driven towards one of the eigenstates of $\hat A(\x)$. Therefore, it is useful to investigate what happens with an ensemble of identically prepared systems. Because of the CSL dynamics, the ensemble will be constituted of a set of different evolved state vectors, each characterized by a different realization of $w(\x,t)$, and the density matrix is the object that best serves to describe the ensemble's evolution.

The density matrix is defined from Eqs. \eqref{CSLQM} and \eqref{CSLprobab} as
\begin{equation}\label{densitymatrix0}
	\hat	\rho(t) \equiv \int_{-\infty}^{\infty} P(w) dw \frac{|\psi , t \ket \bra \psi, t| }{  \bra \psi, t | \psi, t \ket  }
\end{equation}
From Eq. \eqref{densitymatrix0}, it can be shown that the corresponding density matrix evolution satisfies
\begin{equation}\label{densitymatrixevol0}
	\frac{ \partial \hat \rho (t) }{\partial t} = - i [\hat H, \hat \rho (t)] - \frac{\lambda_0}{2} \int d^3x \: [\hat A(\x), [\hat A(\x), \hat \rho(t) ]   ].
\end{equation}
As a consequence, the ensemble expectation value of any observable corresponding to an operator $\hat O$ can be obtained as $\overline{\bra \hat{O} \ket } =$ Tr $\{   \hat O \hat \rho (t) \}$.

Note that, from Eq. \eqref{CSLQM} [or equivalently \eqref{densitymatrixevol0}], one can also identify an effective collapse rate for each particle species $\lambda_i \equiv \lambda_0 ( m_i/m_0)^2$, i.e.  the strength of the collapse scales with the mass of the particle.  This leads to one of the most important features of collapse models:  the \textit{amplification mechanism}.  For microscopic systems (for instance, a single electron) the collapse is very weak, so superpositions can occur. For macroscopic objects (for example, $10^{24}$ atoms) the collapse is very strong, implying a rapid and efficient collapse of the wave function.

If the model is correct, accurate values of $\lambda_0$ and $r_c$ have to be determined by empirical evidence. It should be noted that as the way the model was constructed, relativistic effects have not been taken into account; also, in ordinary matter, the collapse is mostly caused by nucleons. Until experiments can provide precise values of the parameters, one can adopt some estimated values based on reasonable arguments. For example, the choice $r_c \simeq 10^{-5}$ cm, originally suggested by GRW, was based on the assumption that $r_c$ should be large with respect to the atomic dimensions and to the mean spreads around the equilibrium positions of the lattice points of a crystal.  Similar arguments, which involve human perception time and the number of nucleons marking the quantum-classical threshold, lead to the choice  $\lambda_0 \simeq 10^{-16}$ s$^{-1}$ \cite{Pearle76,Ghirardi86,Pearle89,Bassi1,Pearle1994}.

\section{The proposed framework}\label{secCSLposta}

In this section, we present one of the possible extrapolations of the non-relativistic CSL model, described in the previous section, to more general situations in which relativistic effects cannot be ignored. In our view, the proposed model is a natural extension of the original one, since it captures the essence and main features of the \textit{mass proportional} CSL model. Also, it can be applied directly to the cosmological case; and in particular, to the inflationary universe. To accomplish this, we will follow some options discussed in \cite{Bengo20Long}.

The background metric will be given by a spatially flat FLRW spacetime. In conformal comoving coordinates the metric components are $g_{\mu \nu}=a^2(\eta ) \eta_{\mu \nu}$, with $ \eta_{\mu \nu}$ the Minkowski metric with signature $(-,+,+,+)$. We also introduce the Hubble parameter $H$, and then the comoving Hubble parameter is given as $\mH = a'/a = aH$, with primes over functions denoting derivative with respect to conformal time $\eta$. Inflation is characterized by a shrinking comoving Hubble radius $d/d\eta[  \mH^{-1}]<0$, with $H \simeq$ constant. Also, if needed, we can fix the normalization of the scale factor today, so $a_0 = 1$.

We proceed now to characterize the metric perturbations. In particular, we choose to work in the longitudinal gauge where the line element associated to the metric is
\begin{equation}
	ds^2 = a^2(\eta) \left[ - (1+2 \Psi) d\eta^2 + (1-2 E) \delta_{ij} dx^i dx^j \right]
\end{equation}
with $E$ and $\Psi$ being scalar fields corresponding to scalar perturbations at first order. In fact, by assuming that there is no anisotropic stress, Einstein's equations lead to  $\Psi = E$. From now on, we will use this result and refer to $\Psi$ as the Newtonian potential. Moreover, in this gauge, $\Psi$ represents the curvature perturbation (i.e. the intrinsic spatial curvature on hypersurfaces on constant conformal time for a flat universe).

During inflation, the matter degrees of freedom are represented by a single scalar field $\phi(\x,\eta)$, called the inflaton.  Furthermore, we separate $\phi$ in its homogeneous part $\phi_0 (\eta)$ plus a small inhomogeneous perturbation $|\dphi (\x,\eta)| \ll |\phi_0|$. The field $\phi_0$ will be treated in a classical way,  while the perturbation $\dphi$ can be described as a QFT in a flat FLRW background metric.  The condition for an inflationary phase requires that the potential energy $V$ of the scalar field dominates over the kinetic term $\phi'^2 /2a^2$. Furthermore, we will focus on slow roll inflation, where the slow roll parameter is defined as $\epsilon \equiv 1-\mH'/\mH^2 $. During slow roll inflation $\epsilon \ll 1$, and the inflationary phase ends when $\epsilon \simeq 1$.

\subsection{Semiclassical Gravity}

Given the fact that we have not yet at our disposal a complete and satisfactory theory of quantum gravity, we will employ the \textit{semiclassical gravity} (SCG) formulation. This framework describes the way in which the gravitational  and matter degrees of freedom are related to each other. We consider the SCG approach as an effective setting rather than something akin to a fundamental theory. The SCG framework is characterized by Einstein semiclassical equations
\begin{equation}\label{EE}
	R_{ab} - \frac{R}{2} g_{ab} = 8 \pi G \bra \hat T_{ab} \ket
\end{equation}
where $R_{ab}$ and $R$ are the Ricci tensor and scalar respectively; also we are neglecting the contribution of a cosmological constant. The expectation value of the energy-momentum tensor $\bra \hat T_{ab} \ket$, characterizing the matter fields, acts as a source of the spacetime geometry. In this way  the  description of gravity in terms of the metric is always taken as classical.

The inflationary epoch is presumed to develop at energy scales smaller than the Planck mass, therefore, one expects that the SCG approach can be considered as a suitable framework. For our case of interest, the use of SCG possess two main conceptual advantages. First, the spacetime, and thus the metric, is always classical. There is no issue with the ``quantum-to-classical transition'' in the characterization of the spacetime. We will not need to justify the transition from ``metric operators'' (e.g. $\hat \Psi$) to classical metric variables (such as $\Psi$). The fact that the spacetime remains classical is of great importance when including the CSL model. This is because the ``localization'' (or collapse) of the wave function is regarded as a physical process occurring in time. Consequently, it is preferred to have a setting that admits consideration of full spacetime notions \cite{Tejedor12,Pedro18}.  The second advantage is that it allows to present a transparent picture of how the primordial perturbations are born from the wave function collapse: the initial state of the universe (i.e. the one resulting after a few e-folds since the beginning of inflation) is described by the completely symmetric Bunch-Davies (BD) vacuum and the equally symmetric FLRW spacetime; the symmetry being spatial homogeneity and isotropy. Then, after the CSL mechanism has ended, i.e. when the wave function has been effectively localized/collpased, the state emanating from the collapse needs not to share the symmetries of the initial state.  After the collapse, the gravitational degrees of freedom are assumed to be accurately described by Eq. \eqref{EE}. Thus, in this post-collapsed state, in which the symmetries of the original state have been lost, the expectation value $ \bra \hat T_{ab} \ket$ yields a geometry that generically will no longer be homogeneous and isotropic.

\subsection{Main equations of the extended CSL model}

We now  introduce the two main equations that characterize our version of the CSL model. The first is the evolution equation adapted to the cosmological context:
\begin{eqnarray}\label{cslmaster}
	| \Phi , \eta \ket &=&  \hat T \exp \bigg\{   \int_\tau^\eta d \eta \int d^3 x   \sqrt{|g |}  \bigg[  -i  \hat \mH (\x,\eta)  \nn
	&-&   \frac{ 1}{4 \lambda}  \bigg(    W(\x,\eta) - 2 \lambda  \hat C    (\x,\eta)   \bigg)^2     \bigg]  \bigg\}   \nn
	&\times&	| \Phi , \tau \ket
\end{eqnarray}
where $\hat T$ is the time-ordering operator, $d\eta d^3x \sqrt{|g |}$ is the 4-volume associated to the background metric $g_{\mu \nu} $, and $\tau \to -\infty$ is the conformal time at the beginning of inflation. The operator $\hat \mH$ corresponds to the Hamiltonian density of the system. In addition, the CSL parameter $\lambda$ and the collapse generating operator $\hat C$ appear here. We will be more specific about these objects later on. The classical scalar  field $W(\x,\eta)$ is of white noise type,  whose probability is given by the second equation which constitutes the Probability Rule:
\begin{equation}\label{cslprobab}
	P(W) dW =   \bra \Phi , \eta | \Phi, \eta \ket \prod_{\eta'=\tau}^{\eta-d\eta} \frac{| g|^{1/4} d W(\x,\eta')}{\sqrt{2 \pi \lambda/d\eta}}.
\end{equation}

Note that we have chosen the time coordinate to be the conformal time $\eta$, which implies $| g|^{1/2}  = a^4$. The justification of such an election is based on the widely known result  indicating that, in these coordinates,  the ``free'' Hamiltonian  of a scalar field in a flat FLRW spacetime is mathematically equivalent to the one of  a scalar field in Minkowski spacetime with time dependent mass.
Furthermore, from Eqs. \eqref{cslmaster} and \eqref{cslprobab}, and assuming that the initial states are normalized $\bra \Phi, \tau | \Phi, \tau \ket = 1$,  one obtains $\int P(W) dW = 1$ as expected.

The density matrix operator, which is an useful expression describing the ensemble's evolution, is constructed from the state vectors and their associated probabilities, here given by Eqs. \eqref{cslmaster} and \eqref{cslprobab}:
\begin{equation}\label{densitymatrix}
	\hat	\rho(\eta) \equiv \int_{-\infty}^{\infty} P(W) dW \frac{|\Phi , \eta \ket \bra \Phi, \eta | }{  \bra \Phi, \eta | \Phi, \eta \ket  }.
\end{equation}
According to Eq. \eqref{densitymatrix}, the corresponding density matrix evolution is
\begin{equation}\label{densitymatrixevol}
	\frac{ \partial \hat \rho (\eta) }{\partial \eta} = - i [\hat H, \hat \rho (\eta)] - \frac{\lambda a^4}{2} \int d^3x \: [\hat C(\x,\eta), [\hat C(\x,\eta), \hat \rho(\eta) ]   ]
\end{equation}
where $\hat H$ is the total ``free'' Hamiltonian\footnote{Here, we have abused the notation, so keep in mind that $\hat H$ is not related to the Hubble parameter.}, i.e. $\hat H \equiv \int d^3x \sqrt{|g|} \hat \mH$.
As a first step, let us ignore the CSL term in the evolution equations and concentrate on finding  the ``free'' Hamiltonian of the theory during inflation. Given that we are working within the semiclassical gravity framework, only the matter degrees of freedom will be quantized. Consequently, the field variable operator can be taken as $y(\x,\eta)  \equiv  a(\eta)    {\dphi}    (\x,\eta)$. Expanding the action of the standard inflationary model (i.e. a single scalar field--the inflaton--with canonical kinetic term and minimally coupled to gravity) up to second order in the perturbations, one can find the action associated to the matter perturbations. Thus, the second order action is $ S^{(2)}  =  \int d^4x  \mathcal{L}^{(2)}_y $,  where
\begin{eqnarray}\label{action2}
	\mathcal{L}^{(2)}_y  &=& \frac{1}{2} \bigg[  y'^2   - (\nabla y)^2  - y^2 a^2  V_{, \phi \phi} +  \frac{a''}{a} y^2\bigg] \nn
	&+& a [4\phi_0' \Psi' y - 2 a^2 V_{, \phi}  \Psi y  ]
\end{eqnarray}
and $V_{, \phi}$ indicates partial derivative with respect to $\phi$.

The previous Lagrangian can be used to obtain the Hamiltonian density involved in Eq. \eqref{cslmaster}. We define the canonical momentum $p(\x,\eta) \equiv  \partial \mathcal{L}^{(2)}_y/ \partial y = y'$, in this way, the Hamiltonian $\mathcal{H}_y \equiv p y'-    \mathcal{L}^{(2)}_y $ is
\begin{eqnarray}\label{hamilty}
	\mathcal{H}^{(2)}_y  &=& \frac{p^2}{2}     +  \frac{(\nabla y)^2}{2}   + \frac{ y^2}{2}    \bigg( a^2  V_{, \phi \phi} -  \frac{a''}{a} \bigg) \nn
	&-& a^2 H M_P \sqrt{\epsilon}  (  4 \Psi' + 6 a H \Psi   )y
\end{eqnarray}
where we have used $\phi_0' = a H M_P \sqrt{2 \epsilon} $ and the slow roll approximation $-V_{,\phi} \simeq 3 H \phi_0'/a$.

At this point we promote $y$ and $p$ to quantum variables, with equal time commutator
\begin{equation}\label{commutator}
	[\hat y (\x,\eta), \hat p (\textbf{y},\eta)] = i \delta(\x - \textbf{y})
\end{equation}

Also, recall that we are working within the SCG framework, where the metric variables are always classical and only the matter fields are quantized. In particular, in our approach,  because of the CSL mechanism together with SCG, the metric would be changing from a homogeneous and isotropic spacetime to another one with actual inhomogeneities/anisotropies. The latter in fact would become manifested in the metric perturbations such as $\Psi$. As a consequence, the terms involving $\Psi$ in  Hamiltonian \eqref{hamilty}, could be treated as reflecting the effect on $\phi$ of the metric response to the ``fluctuations'' of the field itself $\dphi$.  In other words, the $\Psi$-terms can be considered as a back-reaction on the evolution of the field.  However, in \cite{Leon2020} was shown that these terms are of second order in the slow-roll parameters. Therefore,  the  $\Psi$-terms  in  the Hamiltonian \eqref{hamilty} can be ignored at the leading order considered in this work. Thus, the total ``free'' Hamiltonian during inflation, turns out to be [see the first term in the r.h.s. of Eq. \eqref{densitymatrixevol}]
\begin{eqnarray}\label{csltermino1}
	&   & \hat H =  	 \int d^3 x   \sqrt{|g |}  \hat \mH (\x,\eta) \nn
	& &     \int d^3 x  \left[    \frac{\hat p^2}{2}     +  \frac{(\nabla \hat y)^2}{2}   + \frac{ \hat y^2}{2}    \bigg( a^2  V_{, \phi \phi} -  \frac{a''}{a} \bigg)  \right] . \nn
\end{eqnarray}

\subsection{The collapse rate parameter}\label{sec_lambdaR}

Let us turn our attention to the CSL terms in the evolution equation, Eq. \eqref{cslmaster} [or equivalently Eq. \eqref{densitymatrixevol}]. As mentioned above, the parameter $\lambda$ is the collapse rate. Notice that, in the original GRW collapse model (as well as in the mass proportional CSL model), there is an effective collapse rate that depends on the mass of the particle involved. This feature leads to the \textit{amplification mechanism}, which basically states that as more massive particles are involved, their collapse rates become strengthened. On the other hand,  R. Penrose and L. Diosi have, for a longtime, advocated that  the collapse of the wave function  might be a dynamical process, in which the underlying mechanism might be related to gravitational interaction \cite{Diosi84,Diosi87,Diosi89,Penrose96}. Driven by these ideas, and as extensively discussed  in Ref. \cite{Bengo20Long}, we think  it is natural to consider that the collapse rate should incorporate some aspects tied to the spacetime curvature for systems where this factor is important. A possible extrapolation of $\lambda$ to more general regimes is the one that assumes an explicit dependence of $\lambda$ upon the spacetime curvature. That is, in laboratory experiments where one can safely presume a flat spacetime, the value of the CSL parameter is approximately $\lambda_0$. While in a regime where the spacetime curvature is strong, such a value would completely change.

As we have mentioned in the Introduction (as well as in Ref. \cite{Bengo20Long}), there are several ways to include a spacetime curvature dependence in the collapse rate parameter.  For example, if one were restricted only to include the geometrical features of the spacetime, a reasonable parameterization would be $\lambda(W^2)$, where $W^2 \: \equiv W_{\alpha \beta \mu \nu} W^{\alpha \beta \mu \nu} $ with $W_{\alpha \beta \mu \nu}$  the Weyl tensor.  However, in the case of a FLRW spacetime $W =0$,  hence one would need to include the first order perturbations of the metric so $W \neq 0$. The corresponding analysis, although valid, would require second-order perturbation theory [see e.g. Eq.  \eqref{densitymatrixevol}] which is beyond the scope of the present work.  On the other hand, if it is assumed that the relevant curvature scalar, present in the parameterization of $\lambda$, is that due to the presence of a local matter distribution, then a dependence of scalars constructed using the Riemann tensor seems to be the reasonable option (because of Einstein's field equations).  Once again, there are many choices for expressing $\lambda$ as a function of a curvature scalar constructed using $R^\alpha_{\: \beta \mu \nu}$, e.g. $R$, $R_{\mu \nu} R^{\mu \nu}$,  $R_{\alpha \beta \mu \nu} R^{\alpha \beta \mu \nu}$, etc.\footnote{In Appendix A of Ref. \cite{Bengo20Long} we introduced a toy model in which the CSL parameter $\lambda$ depends on the Kretschmann curvature scalar $ R_{\alpha \beta \mu \nu} R^{\alpha \beta \mu \nu}$ (the case for $\lambda(W^2)$ was also considered). This model, however, did not involve a FLRW spacetime; instead, we analyzed the spacetime corresponding to a static spherical object of certain radius and a homogeneous mass-density distribution.}

In Refs. \cite{Modak14,Modak15}, motivated by finding a resolution to the black hole information loss paradox based on the CSL mechanism, the authors assumed that $\lambda$  is determined by the Ricci scalar $R$ as
\begin{equation}\label{lambda0}
	\lambda = \lambda_0 \left[  1 +    \left(    \frac{R}{\mu}    \right)^{\alpha}     \right],
\end{equation}
with $\lambda_0$ the value of the CSL parameter corresponding to the non-relativistic CSL models, $\alpha>0$ is a free parameter and $\mu$ corresponds to an appropriate physical scale with dimensions of length$^{-2}$.  We point out that, in Refs. \cite{Modak14,Modak15}, the parameter $\alpha$ was constrained to $\alpha \geq 1$. In flat spacetime (or in laboratory experiments perfomed in regions of the spacetime where the  curvature can be safely neglected) Eq. \eqref{lambda0} reduces to $\lambda_0$, i.e. the standard (non-relativistic) CSL parameter.  The advantage of  Eq. \eqref{lambda0}, is that it shows explicitly how the collapse rate is now sensitive to the local curvature.  In other words,  the local curvature scalar $R$ serves to adjust the dominance of the non-linear, stochastic term over the linear term in Eq. \eqref{cslmaster} (or Eq. \eqref{densitymatrixevol} equivalently). Note the former term  is the one responsible for breaking the linear superposition of various basis vectors.  Of course, at the present moment,  one cannot verify or falsify the hypothesis that led to Eq. \eqref{lambda0}. However, in the future,  it is quite possible that direct observations of gravitational effects in the quantum regime could be achieved at laboratory scales,  which,  in turn, will provide strong empirical evidence in favor or against of the proposal that resulted in Eq. \eqref{lambda0}.

Motivated by the aforementioned discussion, we will assume that the collapse rate parameter $\lambda$, used in the CSL model considered in the present paper, is also given by Eq. \eqref{lambda0} identically. In order to provide a reasonable value of $\mu$, we note first that the CSL model, shown in Eq. \eqref{cslmaster}, evolves the state of the quantum field with respect to the conformal time $\eta$. Therefore, we are explicitly considering a particular foliation of the spacetime, in which  the states evolve from one space-like surface of constant $\eta$ to another one.  Additionally, in our proposal,  the rate of collapse  is enhanced by the curvature of the spacetime, according to Eq. \eqref{lambda0}. Hence, in principle, the collapse rate depends on the time parameter defined by the foliation $\lambda = \lambda(\eta)$.

Assuming that the matter content in the universe can be described by the energy-momentum tensor of a perfect fluid $T_{\mu \nu} = (\rho + P) u_\mu u_\nu + P g_{\mu \nu}$, where $\rho$,  $P$ correspond to the energy density and pressure in the rest frame of the fluid respectively, and $u^\mu$ represents its 4-velocity (relative to the observer).  Then,  from Einstein equations (in trace form) $R = -T/M_P^2$, together with equation of state $P/\rho=\omega \simeq$ constant, one obtains
\begin{equation}\label{Riccicosmo}
	R = \frac{\rho} {M_P^2} (1- 3 \omega).
\end{equation}

At this point,  we can discuss the  specific  physical scale $\mu$ employed in our case of interest. During slow roll inflation, $\omega_{\text{inf}} \simeq -1$ and $\rho_{\text{inf}} = 3 M_P^2 \Hi^2 \simeq$ constant, so $R \simeq 12 \Hi^2$.  Let us introduce a spatial physical length of size $\ell_{\text{phys}}= a |\tau|$. This scale\footnote{We note that $|\tau|$ is, to a very good degree of approximation, the size of the comoving particle horizon at any time $\eta$ after the end of inflation. The particle horizon at time $\eta$ can be envisaged as the intersection of the past light cone of an observer at point $p$ with the spacelike surface at $\eta = \tau$. Causal influences must originate inside this region. Only comoving particles whose worldlines intersect the past light cone of $p$ can send a signal to an observer at $p$. Therefore, the physical particle horizon $a|\tau|$ is a good candidate for a physical length to be compared with $R$ during inflation. Conversely, other natural choices do not enhance the term $R/\mu$ during inflation,  e.g. assuming  $\mu = \ell_P^{-2}$, where $\ell_P$ is Planck's length, implies $R/\mu \ll 1$ because inflation involves energy scales less than Planck scale. So, the term $R/\mu$ does not affect the collapse rate, i.e. $\lambda \simeq \lambda_0$.  Evidently, a similar situation occurs if one chooses $\mu = \Hi^2$,  which is essentially $R$ during inflation, hence $R/\mu \simeq 1$,  so $\lambda \simeq \lambda_0$. That is, once again the curvature term does not dominate in Eq. \eqref{lambda0}.} induces the particular choice $\mu = \ell_{\text{phys}}^{-2} $. In this case, we can estimate $|\tau| \simeq 1/(a_\tau  \Hi)$, and  by taking into account the  number of e-folds $N$ since the beginning of inflation $a = e^{N} a_\tau$,  we conclude that  $R/\mu \simeq e^{2N} \gg1 $. Therefore, the choice $\mu = \ell_{\text{phys}}^{-2} $  ensures that, during the full inflationary regime, the collapse rate $\lambda$ in Eq. \eqref{lambda0}, is enhanced by the term $R/\mu$.  In particular, one has
\begin{equation}\label{lambda2}
	\lambda(\eta) \simeq \lambda_0 e^{2 N(\eta) \alpha}
\end{equation}
and from now on we will use this expression for the collapse rate when analyzing the CSL model during inflation.

It is also relevant to analyze the behavior of $\lambda$ in the subsequent cosmic stages after inflation. For pure radiation $\omega_{\text{rad}} =  1/3$,  one obtains that  $R^{\text{rad}} $ vanishes exactly.  For pressure-less matter (dust) $\omega \simeq 0 $, the Ricci scalar is  $R^{\text{matt}} \simeq   \rho^{\text{matt}}/M_P^2$.  After inflation ends, we can approximate the matter content in the universe  by a two-component mixture of radiation and pressure-less matter. The Ricci scalar is then $R = \sum_i  \rho_i  (1- 3 \omega_i)/M_P^2 = R^{\text{matt}}$, where the index $i$ labels each component of the matter content, $i=$ $\{$rad, matt$\}$.  Therefore, after the inflationary regime, the evolution\footnote{The total energy density $\rho \equiv \rho_{\text{matt}} + \rho_{\text{rad}}$ evolves as
	\begin{equation*}
		\rho = \frac{\rho_{\text{eq}}}{2} \left[   \left( \frac{a_{\text{eq}}}{a} \right)^3     + \left( \frac{a_{\text{eq}}}{a} \right)^4 \right].
	\end{equation*}
} of the  quantity $R/\mu$ is given as
\begin{equation}\label{Rmumatt}
	\frac{R}{\mu} 	= \frac{ 3 \rho_{\text{eq}} a_{\text{eq}}^2}{2 \rho_{\text{inf}} a_\tau^2 } \frac{a_{\text{eq}}}{a}
\end{equation}
where we used the choice $\mu = \ell_{\text{phys}}^{-2}$.

The quantity $a_{\text{eq}}$ denotes the scale factor evaluated at the matter-radiation equality epoch.  In particular, one has $a_{\text{eq}} = e^{N_{\text{eq}}} a_\tau$, where $N_{\text{eq}}$ is the total number of e-folds from the beginning of inflation up to $a_{\text{eq}}$. Moreover, assuming an inflationary energy scale of $\rho_{\text{inf}}^{1/4} \simeq 10^{15}$ GeV and that the energy scale at the radiation-matter equality epoch is   $\simeq 1$ eV, we have $\rho_{\text{eq}} \simeq 10^{-96} \rho_{\text{inf}}$. Also, if we consider that inflation lasts approximately $60$ e-folds and that $a_{\text{eq}} \simeq 10^{-4} a_0$,  then  $N_{\text{eq}} \simeq 115$. Taking into account all these estimates, Eq. \eqref{Rmumatt}  yields
\begin{equation}\label{Rmumatt2}
	\frac{R}{\mu}  \simeq  \frac{a_0}{a}.
\end{equation}
The above equation implies that at the onset  of the matter-radiation equality era $R^{\text{eq}}/\mu  \simeq 10^4$, and then it decays\footnote{ In the radiation dominated epoch, $a < a_{\text{eq}}$,  the quantity  $R/\mu$ also decays as $a$. However, taking e.g. the primordial nucleosynthesis epoch as the beginning of the radiation dominated epoch, one obtains $R/\mu \simeq 10^9$, so from Eq. \eqref{lambda0} we find that the CSL parameter is constrained between $\lambda_0 10^{9 \alpha} \geq \lambda (\eta) \geq \lambda_0 10^{4 \alpha}$.
	Assuming the allowed experimental range $10^{-13}$s$^{-1}$ $\geq \lambda_0 \geq 10^{-19}$ s$^{-1}$ and $\alpha \leq 2$, we find that the parameter $\lambda (\eta)$ is consistent with the bound $\lambda_{\text{RD}} \leq 10^{5}$s$^{-1}$, which was obtained in the radiation dominated epoch. That value was found in Ref.  \cite{BassiCMB} using the mass proportional non-relativistic CSL model and spectral distortions in the CMB.
} as $a^{-1}$. Specifically, assuming a matter dominated regime,  the quantity  $R/\mu$ evaluated today is approximately  of order $10^0$. Consequently, in the matter dominated epoch, $a > a_{\text{eq}}$,  from expression \eqref{lambda0},  we find that  the CSL parameter $\lambda$ will be constrained between
\begin{equation}\label{lambdamatt}
	10^{4 \alpha} \lambda_0 \geq \lambda(\eta) \geq  \lambda_0.
\end{equation}

The experimental constraints on $\lambda_0$ are between \cite{Bassiexp21}
$$10^{-10}   \textrm{s}^{-1} \geq \lambda_0 \geq 10^{-19} \textrm{s}^{-1}.$$
Thus, for $\alpha \leq 2$ the variation of $\lambda$ within the interval \eqref{lambdamatt} due to the matter dominated cosmological epoch, can fit the allowed experimental range. In particular, if we consider  values close to the lowest empirical bounds, e.g. $ \lambda_0 = 10^{-19}$s$^{-1}$.

The previous analysis serves to illustrate that, as a result of the choice $\mu = \ell_{\text{phys}}^{-2} $,  the term $R/\mu$ becomes very dominant in Eq. \eqref{lambda0} during inflation. However,  after the end of inflation $R/\mu$ decays as $a^{-1}$. In particular, in the matter dominated epoch, the CSL rate parameter is essentially the same as the non-relativistic version of the CSL model.   This is a desirable feature to be included in our model because  all  results and observational bounds, regarding the CSL model, obtained from contemplating this cosmic stage,  remain intact \cite{BassiCMB,Corral20,Nucamendi20}.

\subsection{The collapse generating operator}\label{sec_smearing}

The next important element for the CSL cosmological model is the collapse generating operator.  As we have argued in the Introduction, we are interested in a collapse operator that is obtained from covariant objects that reduce, in the simple non-relativistic regimes encountered in laboratory situations, to the ``mass density'' or the ``energy density''.  Specifically, we would like to recover something that can be interpreted as the mass density operator used in the mass proportional CSL model [see Eq. \eqref{Acolop}].  One possible covariant object, as mentioned in \cite{Bengo20Long}, is the scalar $\Gamma \equiv  (T_{\mu \nu}T^{\mu \nu} )^{1/2}$.  Moreover, given that our main focus is to implement the CSL model during inflation, and obtain the primordial spectrum, we will be working within the framework of cosmological perturbation theory at linear order. As a consequence, our choice for the collapse operator $C$ will be motivated by the first order perturbations associated to the scalar $\Gamma$. In particular, the collapse operator will be derived from:
\begin{equation}\label{colop}
	C(\x,\eta) \equiv  \frac{1}{M_P}\int d^3 y \sqrt{|h |}  s(\x,\y;\eta)   \delta \Gamma(\y,\eta)
\end{equation}
where,
$d^3y \sqrt{|h |}$  is the 3-volume associated  to the spatial metric $h_{ij} $ ,  the integration is over one of the hypersurfaces of constant time $\eta$, and
\begin{equation}\label{deltagamma}
	\delta \Gamma =  \frac{1}{ 2 ( \bar T_{\mu \nu}  \bar T^{\mu \nu} )^{1/2}   }  (   \bar T^{\mu \nu} \delta T_{\mu \nu}    +  \bar T_{\mu \nu} \delta T^{\mu \nu}    )
\end{equation}
with $\bar T_{\mu \nu}$ denoting the background part.  The function $s(\x,\y;\eta)$ is a smearing function  defined over the spatial hypersurfaces of constant $\eta$ with dimensions of length$^{-3/2}$ (as it occurs in the non-relativistic CSL model). The factor $M_P^{-1}$ in Eq. \eqref{colop}, fixes the correct dimensions of $C$ and can also be considered  to play a similar role as the reference mass usually introduced in non-relativistic CSL models (in which the mass of a nucleon is commonly used).

With the same spirit as the original CSL model, where the smearing function helps to localize a macroscopic body in position space, here we adopt the view that $s(\x,\y;\eta)$,  induces a ``localization'' of the fields associated to $\delta \Gamma$. In other words,  at first sight $s(\x,\y;\eta)$ should depend on the ``quantum uncertainties'' of the field variables rather than the uncertainties of the position operator as in the non-relativistic CSL model.

In the inflationary case, one can find the explicit dependence on the field variables for the operator $\delta \hat \Gamma$. Using Eq. \eqref{deltagamma}, and taking into account  the canonical quantum variables $\hat y, \hat p$, while neglecting the terms proportional to $\Psi$ [see discussion after Eq. \eqref{commutator}],  one has
\begin{equation}\label{deltagamma3}
	\delta \hat  \Gamma  = H M_P\sqrt{2 \epsilon}  \left(     \frac{\hat p}{a^2} + 5 H \frac{\hat y}{a}     \right)
\end{equation}

In the non-relativistic CSL model, the smearing function introduces the second parameter of the model, namely the smearing length $r_c$. The parameter $r_c$ is a length scale characterizing the localization of the wave function through the collapse mechanism. For a single particle, if the width of its wave function is much less than $r_c$, the effect of the collapse becomes negligible. On the contrary, if the width is larger than $r_c$ the collapse effects become significant. Therefore, as we have argued, the quantum uncertainties of the position operator, associated to the width of the wave function in position space, play an important role in defining the smearing function and in characterizing the parameter $r_c$.

In our view, it is not at all obvious that there is a simple one-on-one connection between the values of $r_c$, characterizing laboratory experiments effectively described by non-relativistic Quantum Mechanics, and the parameters of a CSL-like model describing  a complete different regime such as the inflationary universe \cite{Bengo20Letter,Bengo20Long}. As a matter of fact,  in this regime certain notions are lost, like the position operator of a particle; but, more fundamental ones emerge, e.g. the quantum fields.

In order to move forward, we will adopt as fundamental the relation between ``quantum uncertainties'' and the smearing function.  Intuitively, we will think the smearing function $s$ as strongly dependent on the ``quantum uncertainties'' of the system under consideration. For example, if the system can be effectively described by non-relativistic Quantum Mechanics, then ``quantum uncertainties'' in the position operator should serve to define the size of the smearing region, and thus set the value of $r_c$. On the other hand, if the system we want to analyze is, e.g. the radiation emitted by Black Holes, then $s$ should depend on the ``quantum uncertainties'' of the matter fields and, quite possibly, on the background spacetime curvature.  However, it is a well-known fact that, strictly speaking, the uncertainty of a quantum field is not well-defined generically because vacuum expectation values like $\bra \hat \varphi^2 (x) \ket_{\text{vac.}}$ diverge.  In order to take the next step,  we can focus on the two-point correlation function(s). In fact, until a full covariant CSL theory is formulated, we can only make progress by considering particular smearing functions with the desired characteristics.

Returning to our case of interest, the operator 	$\delta \hat  \Gamma $ in Eq. \eqref{deltagamma3} depends linearly on the field variables $\hat{p}/a^2$ and $\hat y /a$.  Without loss of generality,  we will analyze the 2-point function of $\hat y/a $, and the corresponding analysis of $\hat{p}/a^2$ will proceed in an equivalent manner. Let us recall that $\hat{y} /a =  \hat \dphi$, and concentrate on very small proper wavelengths (high-frequencies) $\lambda_{\text{phys}} \ll H^{-1}$.  In the vacuum state, characterized by the standard Bunch-Davies vacuum   \cite{Birrell}, the 2-point function of $\hat{y} /a $ is
\begin{equation}\label{2puntosdentro}
	\bra  \hat 	\dphi  (\x_1,\eta)   	\hat \dphi (\x_2,\eta) \ket_{\text{BD}} \propto \frac{1}{a^2 r^2}
\end{equation}
where $r \equiv | \x_1 - \x_2 |$. On the opposite regime,   $\lambda_{\text{phys}} \gg  H^{-1}$, the 2-point function in the BD vacuum becomes
\begin{equation}\label{2puntosfuera}
	\bra  \hat 	\dphi  (\x_1,\eta)   	\hat \dphi (\x_2,\eta) \ket_{\text{BD}}\propto H^2
\end{equation}

Therefore, as inflation takes place, the 2-point function decreases  because the physical distance ($ar$) increases. But, as $ar$ becomes larger than $H^{-1}$, the 2-point function remains basically a constant of order $H^{2}$. Thus, we can make use of $H$ as characterizing the size of the smearing region for the field variable $\hat y/a$ whose ``uncertainty'' the CSL model will attempt to decrease. In view of the previous discussion and   the analysis of the 2-point function, we propose the following smearing function:
\begin{equation}\label{smearing}
	s(\x,\y;\eta) \equiv \frac{H^{3/2}}{(2 \pi )^{3/2}} e^{- \frac{1}{2} H^2 a(\eta)^2 |\x-\y|^2}
\end{equation}
Here, $H$ acts akin to the $r_c$ parameter of the non-relativistic CSL model.  To further illustrate this point, we can write the explicit form of the collapse operator obtained from Eq. \eqref{colop} by gathering all its elements. Using the smearing function in Eq. \eqref{smearing},  the collapse operator  is then
\begin{equation}\label{colop2}
	\hat{C} (\x,\eta) = \frac{a^3 H^{3/2}}{(2 \pi )^{3/2} M_P}  \int d^3 y \:  e^{-\frac{1}{2} a^2 H^2r^2} \delta \hat \Gamma (\y,\eta)
\end{equation}
Switching to Fourier space,  and using Eq. \eqref{deltagamma3}, the collapse operator can be expressed as:
\begin{equation}\label{colopfourier}
	\hat{C}_{\nk} (\eta)=\left(\frac{2 \epsilon}{H} \right)^{1/2}    e^{-\frac{k^2}{2 a^2 H^2} } \left(     \frac{\hat p_{\nk}}{a^2} + 5 H \frac{\hat y_{\nk}}{a}                 \right)
\end{equation}
We observe that, for modes such that $k \gg a H$, the operator $	\hat{C}_{\nk} (\eta)$ becomes exponentially suppressed. That is, the effect of the CSL term on the evolution of the quantum state is negligible. On the contrary, for modes such that $k \ll a H$, the effect of the collapse (induced by the CSL evolution), grows stronger.  This is the reason  why in our proposed model $H$ acts in a similar manner as the $r_c$ parameter.  Furthermore, this is consistent with the fact that super-Hubble modes ($k \ll a H$) contribute the most to the observed spectrum. Namely, we expect the collapse affects the most to the state characterizing the modes which have the strongest presence in the primordial inhomogeneities, and which later become imprinted in the CMB. For a more precise analysis regarding this subject, we invite the reader to check  \ref{AppB}.

After presenting our version of the CSL model, and how it is adapted to inflation, in the next section, we will focus on obtaining one of the main predictions of the inflationary paradigm: the primordial power spectrum. However, before ending this section, let us discuss how the modified Schr\"odinger equation, corresponding to the mass proportional CSL model employed in ordinary laboratory circumstances, can be recovered from our proposal.

The generalized collapse operator, $C(\x,\eta)$, as given by Eq. \eqref{colop},  involves three distinctive elements: (i) the perturbation associated to the quantity $\Gamma$, (ii) the smearing function $s$ and (iii) the reference mass $M_P$.  Regarding (i), given the definition $\Gamma \equiv  (T_{\mu \nu}T^{\mu \nu} )^{1/2}$, it is clear that in the non-relativistic limit, $\Gamma \to \rho$ where $\rho$ denotes the mass density function.\footnote{This case is analogous to that encountered in General Relativity when one is interested in recovering the Poisson equation for the Newtonian potential from the Einstein field equations, $\nabla^2 \Phi = 4 \pi G \rho $ . The matter density source $\rho$ comes from the energy-momentum tensor} In particular, for several non-relativistic massive particles $\rho(\x) = \sum_i m_i N_i(\x)$,  where $N_i(\x)$ is the particle number density for a particle of type $i$ at $\x$ and $m_i$ is the mass of the corresponding particle species. With respect to element (ii),  from the arguments presented in this section [recall discussion after Eq. \eqref{deltagamma3}], in the regime where non-relativistic Quantum Mechanics can be used, the smearing function takes the form $s \to (\pi r_c^2)^{-3/4} \: e^{-|\x-\y |^2/2 r_c^2} \:$. And as regards to (iii), the reference mass suitable at laboratory scales corresponds to the mass of a nucleon $m_0$ instead of the reduced Planck mass $M_P$.  Furthermore, since ordinary experiments are performed in regions of the spacetime where the local curvature can be ignored, from Eq. \eqref{lambda0}, we have that $\lambda(R) \to \lambda_0$.  In view of the previous arguments, in the non-relativistic quantum regime, the CSL term proposed in Eq. \eqref{cslmaster}, which modifies the Schr\"odinger equation, reduces to
\begin{eqnarray}\label{csllaboratorio}
 & & \int_\tau^\eta d \eta \int d^3 x   \sqrt{|g |}    \frac{ 1}{4 \lambda(R)}  \bigg(    W(\x,\eta) - 2 \lambda(R)  \hat C    (\x,\eta)   \bigg)^2     \nn
&\to & \int_{t_i}^t d t \int d^3 x    \frac{ 1}{4 \lambda_0}  \bigg(    w(\x,t) - 2 \lambda_0  \hat A    (\x,t)   \bigg)^2 ,
\end{eqnarray}
where $\hat A(\x,t)$ is given in Eq. \eqref{Acolop}.  Note that we have deliberately made a distinction between $w$ and $W$ due to the fact that there is very little knowledge about the physical origin of this term, so, in principle we cannot assure that the noise term is the same for all physical systems and for all scales (however its statistical features are the same).

\section{The Newtonian potential and an equivalent power spectrum}
\label{secPk}

In this section we present the derivation of an important equation, relating the primordial curvature perturbation and the proposed CSL model of the previous section. Also, we provide an expression for the primordial power spectrum, which is equivalent but not exactly equal to the standard one.

The perturbed Einstein equations (EE) at linear order  in the longitudinal gauge can be combined into \cite{mukhanov92,mukhanov2005}:
\begin{equation}\label{EE1}
	\nabla^2 u = z \left( \frac{v}{z}  \right)' ,  \qquad    v = \theta \left( \frac{u}{\theta}  \right)'
\end{equation}
with the following definitions:
\begin{subequations}\label{defEE1}
	\begin{equation}
		v \equiv a \left(   \dphi + \frac{\phi_0'}{\mH} \Psi       \right), \qquad u \equiv \frac{2 M_P^2 a \Psi}{\phi_0'}
	\end{equation}
	\begin{equation}
		z \equiv \frac{a \phi_0'}{\mH}, \qquad \theta \equiv \frac{1}{z}.
	\end{equation}
\end{subequations}

By using the definition of the slow roll parameter $\epsilon $ and $z= a  M_P \sqrt{2 \epsilon} $, we can combine equations \eqref{EE1} to obtain
\begin{equation}\label{masterpsi}
	\Psi + \mH^{-1}   \Psi' = \sqrt{ \frac{\epsilon}{2}}  \frac{ \bra \hat y \ket}{a  M_P}
\end{equation}
where we have made use of the SCG approach. Also, Eq. \eqref{masterpsi} is exact, i.e no approximations were made. This equation relates the quantum expectation value of the matter degrees of freedom, obtained from the CSL collapse mechanism, and the primordial curvature perturbation represented by the Newtonian potential, which is always classical. In particular, in the vacuum state $ \bra \hat y \ket_{\text{vac}} = 0$,  meaning that no curvature perturbations are present. It is only after the collapse has taken place that $ \bra \hat y \ket \neq 0$, and the primordial perturbation is ``born.''

We introduce a well-known quantity, defined generically as
\begin{equation}\label{relacionRyPsiexacta}
	\mR \equiv \Psi  + \left(  \frac{2 \rho}{3}   \right) \left( \frac{\mH^{-1}  \Psi' + \Psi}{\rho + P}\right)
\end{equation}
where $\rho$ and $P$ are associated to the type of matter driving the expansion of the universe.  A main feature of the quantity $\mR$ is that, for adiabatic perturbations, it is conserved for super-Hubble scales, irrespective of the cosmological epoch one is considering \cite{mukhanov92}.

From the components of the energy-momentum tensor of a single scalar field, we have $\rho + P = \phi_0'^2/a^2 = M_P^2 \mH^2 2 \epsilon /a^2$, and because of Friedmann's equation $\mH^2 = a^2 \rho/ 3 M_P^2$, one obtains
\begin{equation}
	\mR = \Psi \left(    1 + \frac{1}{\epsilon}     \right) + \frac{ \mH^{-1} }{\epsilon}    \Psi'
\end{equation}
The above equation is exact, but during slow roll inflation $\epsilon \ll 1$; consequently,  at the lowest order in $\epsilon$,  we can approximate the latter expression as
\begin{equation}\label{R}
	\mR \simeq   \frac{1}{\epsilon}     \left(    \Psi     +  \mH^{-1}  \Psi' \right)  =      \frac{ \bra \hat y \ket}{a M_P \sqrt{2 \epsilon}}
\end{equation}
where in the last equality we have used our main equation \eqref{masterpsi}.

Another important aspect of the quantity $\mR$ is that, in the comoving gauge, it represents the curvature perturbation. In fact, the primordial power spectrum usually shown in the literature is associated to $\mR$. The scalar power spectrum (associated to the curvature perturbation in the comoving gauge and in Fourier space) is defined as
\begin{equation}\label{PSdef}
	\overline{\mR_{\nk}\mR^*_{\nq}} \equiv \frac{2 \pi^2}{k^3} \mP_{s} (k) \delta(\nk-\nq)
\end{equation}
where $\mP_{s} (k)$ is the dimensionless power spectrum. The bar appearing in \eqref{PSdef} denotes an ensemble average over possible realizations of the stochastic field $\mR_{\nk}$. In the CSL inflationary model, each realization will be associated to a particular realization of the stochastic process characterizing the collapse mechanism, which in turn is related to a specific realization of the noise $W$.

On the other hand, our main equation \eqref{masterpsi}, was obtained in the longitudinal gauge. Fortunately, Eq. \eqref{relacionRyPsiexacta} relates $\Psi$, $\Psi'$ and $\mR$  exactly. That is, we can compute the curvature perturbation in the longitudinal gauge (where the SCG framework together with the CSL collapse mechanism generate  the primordial inhomogeneities as given in Eq. \eqref{masterpsi}), and then, we can switch to the comoving gauge in order to compare the primordial spectrum obtained in our model with the standard one. In particular,  we can use approximation \eqref{R} to compute the scalar power spectrum, associated to $\mR_{\nk}$, that results from our main equation \eqref{masterpsi}. This is,
\begin{equation}\label{PSR}
	\overline{\mR_\nk \mR_\nq^*} =  \frac{1}{2 M_P^2 a^2 \epsilon }   \overline{\bra \hat y_\nk \ket \bra \hat y_\nq \ket^*}
\end{equation}

Finally, from definition \eqref{PSdef} and Eq. \eqref{PSR},  we can identify an equivalent scalar power spectrum as:
\begin{equation}\label{masterPS}
	\mP_{s} (k) \delta(\nk-\nq) = \frac{k^3}{4 \pi^2 M_P^2 a^2 \epsilon }  \overline{\bra \hat y_\nk \ket \bra \hat y_\nq \ket^*}
\end{equation}

\section{Analysis of the scalar power spectrum}
\label{secAnalysis}

The next step is to use the CSL model described in section \ref{secCSLposta}, to compute the expectation values in Eq. \eqref{masterPS}. Furthermore,  one needs to evaluate $	\mP_{s} (k) $ in the super-Hubble regime $-k\eta \to 0$ to be able to compare the  corresponding   theoretical prediction with the observational data. The calculations are long but straightforward; all the technical details employed are presented in  \ref{AppA}.

At the lowest order in the slow roll parameter $\epsilon$, the primordial power spectrum obtained is
\begin{eqnarray}\label{PS0}
	\mP_{s}(k)  &=& \frac{H^2}{8 \pi^2 M_P^2 \epsilon} \bigg\{   1 -  \frac{2  \sigma_k }{9}  \bigg[    21 + 29 \gamma_\varepsilon  -58 N  \nn
	&+&   58 \ln (- k \tau)   \bigg]   - F(\sigma_k,-k \eta)|_{-k \eta \to 0} \bigg\}
\end{eqnarray}
where $N$ is the total number of e-folds since the beginning of inflation, and  $\gamma_\varepsilon$ is the Euler-Mascheroni constant. We also introduce  the coefficient
\begin{equation}\label{defsigmak}
	\sigma_k \equiv  k^3 |\tau|^3   \frac{\lambda_0 \epsilon}{H}
\end{equation}
and the function $F(\sigma_k,-k \eta)|_{-k \eta \to 0} $ is defined in \eqref{defF} [see also Eq. \eqref{ReAk2}].

One of the main assumptions used to obtain the primordial spectrum \eqref{PS0} was to fix the free parameter $\alpha = 3/2$, see Eq. \eqref{lambda2}. This is consistent with the previous analysis, in which for $\alpha \leq 2$ and the matter dominated cosmological epoch, the parameter $\lambda$ can fit the experimental bounds coming from the non-relativistic CSL model.  We have analyzed other values  of the free parameter, for example $\alpha = 1,2$ , and we found that these values do not yield the correct shape of the spectrum, i.e. the one favored by observational data.

We proceed to examine the shape and amplitude of the predicted power spectrum \eqref{PS0}. As shown in \ref{AppA},  in the super-Hubble regime,  the function $F$ appearing in the expression for the power spectrum behaves as
\begin{equation}\label{desigualdadchida}
	F(\sigma_k, -k \eta)|_{-k \eta \to 0} \simeq (-k \eta)^6 \to 0
\end{equation}
which means that $F(\sigma_k, -k\eta)|_{-k\eta \to 0}  \ll 1$. The latter result implies that, in Eq. \eqref{PS0}, the  last term of  $\mP_{s}(k)$ can be neglected.

Additionally,  we note that it is possible to express $F(\sigma_k, -k\eta)$ as
\begin{equation}\label{Falt}
	F(\sigma_k, -k\eta) =  \frac{\overline{ \bra \hat  y_\nk^2 \ket  } - 	\overline{ \bra \hat  y_\nk \ket^2  }}{  \overline{ \bra \hat  y_\nk^2 \ket}_{\text{BD}}    }.
\end{equation}
Hence,  the fact that $F(\sigma_k, -k\eta) \ll 1$, when $-k\eta \to 0 $, means that the width of the wave functional associated to $y_\nk$, which in the BD vacuum state is a Gaussian centered at zero, is decreasing during inflation due to the CSL mechanism. In other words, the CSL process is effectively localizing (or ``collapsing'') the initial wave function.

On the contrary, if we turn off the collapse mechanism (i.e. $\lambda=0$), then $\sigma_k = 0$ and $F(0, -k \eta)= 1$ [see also Eq. \eqref{ReAk2}].  Thus, in the absence of the CSL collapse term,  $\mP_{s}(k) =0$. This is an expected result because  if there is no collapse, then the quantum state of  the field remains homogeneous/isotropic, and, as a consequence, there are no inhomogeneities/anisotropies in the spacetime either.

The expression  for $\mP_{s}(k)$ was obtained working at the lowest order in the slow roll parameter $\epsilon$. Henceforth, at this order, the predicted spectrum must be nearly scale invariant for it to be consistent with the observational data. At the next leading order in $\epsilon$, one could find the standard scale dependence reflected in the so called \textit{scalar spectral index}. It is thus clear that any significant departure from a perfect scale invariant spectrum in Eq. \eqref{PS0} would mean a fatal failure for the  CSL model proposed in this work.  We have argued that the last term in Eq. \eqref{PS0} can be neglected if $\lambda \neq 0$. Therefore, the other term that might induce a strong departure from scale invariance is the one with the coefficient $\sigma_k$. Let us focus on this term.

The observed amplitude of the temperature aniso-tropies in the CMB constrains the value of $H$ during inflation, and this value is approximately $H \simeq 10^{-5} M_P \epsilon^{1/2}$.  Moreover, we can choose a particular value of $\lambda_0$ consistent with laboratory experiments, for instance   $\lambda_0 \simeq 10^{-17}$s$^{-1}$, or in Planck units $ \lambda_0 \simeq 10^{-61} M_P $.  From the definition of $\sigma_k$, Eq. \eqref{defsigmak},  it then follows the  estimate
\begin{equation}\label{sigmak1}
	\sigma_k \simeq (k |\tau|)^3   \epsilon^{1/2} 10^{-56}
\end{equation}

Considering  a typical energy scale for inflation  $V^{1/4} \simeq 10^{-3} M_P$, implies $\epsilon \simeq 10^{-3}$. Assuming also  a total duration for inflation of $N=65$, yields $|\tau| \simeq 10^{5}$ Mpc.  In addition, by taking into account $k$ in the range of observational interest \cite{Planck18b}, one has $k \in [10^{-4},10^{-1}]$ Mpc$^{-1}$. Putting together all these values, we find that the coefficient is within the following estimated range:
\begin{equation}\label{sigmak2}
	\sigma_k \in [10^{-55},10^{-46}]
\end{equation}
Consequently, $\sigma_k \ll 1 $ for the range of $k$ that contributes the most to the observed spectrum. We point out that other values of $\lambda_0$, within the  window allowed by laboratory experiments, also lead to a strong suppression of the coefficient $\sigma_k$.

The former analysis indicates that the primordial power spectrum \eqref{PS0} is scale invariant to a very good degree of approximation. Thus, we have found that, at the lowest order in the slow roll parameter, the predicted spectrum is essentially the same as the traditional one, including its amplitude, i.e.
\begin{equation}\label{PS1}
	\mP_{s}(k) \simeq \frac{H^2}{8 \pi^2 M_P^2 \epsilon}
\end{equation}

\section{Conclusions}
\label{conclusions}

The adequate implementation of the CSL model during inflation has been critically analyzed in recent works \cite{MartinShadow,Bengo20Long,Bengo20Letter,Martin20R,Bassi21,Martin21}.  In particular, in Ref. \cite{Bengo20Long} the most important elements to be considered were discussed, according to our point of view. These are: (i) a suitable framework for treating the quantum degrees of freedom and gravity, which we think is provided by the semiclassical gravity approximation, (ii) a collapse-generating operator that can be constructed from covariant objects and, at the same time, can be interpreted as the ``mass density'' operator usually encountered in the laboratory applications of the CSL model, and (iii) a generalization of the CSL parameters that includes a dependence on the spacetime curvature. In this work, we have provided a concrete realization based on these three hypotheses. We have shown that the proposed model yields a primordial scalar power spectrum with the characteristics required to be consistent with observational CMB data, i.e. we have found a nearly scale invariant power spectrum. Moreover, there is no conflict between our proposal and empirical constraints, found in laboratory experiments, on the parameters of the non-relativistic version of the CSL model.

On the other hand,  given that our model is not fully covariant, some potential problems might arise. However, we have taken a first step in exploring the theoretical landscape  corresponding to the extrapolation of the CSL model suitable to more complex systems, in which  gravity and quantum fields play an important role.  Evidently, open issues remain to be analyzed, and we hope to contribute in addressing some of them in the future ahead.

\begin{acknowledgements}
The authors thank the anonymous referees for their comments and suggestions.
G.L. is supported by CONICET (Argentina), and also acknowledges support from the following project grants: Universidad Nacional de La Plata I+D G175 and PIP 112-2020-0100729CO of CONICET (Argentina). G.R.B. is supported by CONICET (Argentina) and he acknowledges support from grant: PIP 112-2017-0100220CO of CONICET (Argentina). We thank Philip Pearle for encouraging this study. We are also especially grateful to Daniel Sudarsky for his constructive comments and suggestions.
	
\end{acknowledgements}

\appendix

\section{Computation of the primordial spectrum}\label{AppA}

The standard primordial power spectrum is expressed in Fourier space. So, we start by defining the Fourier transform of any scalar field as:
\begin{equation}
	f(\x,\eta) = \frac{1}{(2 \pi)^{3/2}} \int_{\mathbb{R}^3}  d^3 k \: f_{\nk} (\eta) e^{i \nk \cdot \x}
\end{equation}

In Fourier space, Eq. \eqref{csltermino1} takes the form
\begin{eqnarray}\label{csltermino1b}
	& & 	 \int_{\mathbb{R}^{3+}} d^3 k  \quad a^4 \hat  \mH_\nk  =  \nn
	& &	\int_{\mathbb{R}^{3+}} d^3 k \quad  \left[  \hat p^*_\nk \hat  p_\nk  + \hat   y^*_\nk  \hat  y_\nk \left( k^2 - \frac{a''}{a}  + a^2  V_{, \phi \phi}    \right)   \right].  \nn
\end{eqnarray}

The quantization procedure will be   carried out  in the Schr\"odinger picture, so it is more convenient to work with real variables, which later can be associated to Hermitian operators. In this way,  we separate the canonical variables in their real and imaginary parts. In Fourier space, this is given as
\begin{equation}
	\hat 	y_\nk \equiv \frac{1}{\sqrt{2}} (\hat y_\nk^\text{R} + i \hat  y_\nk^\text{I}  ), \qquad \hat  p_\nk \equiv \frac{1}{\sqrt{2}} (\hat  p_\nk^\text{R} + i \hat  p_\nk^\text{I}  ).
\end{equation}
The quantum commutator in Eq. \eqref{commutator} implies
\begin{equation}
	[\hat y_\nk^s , \hat p_\nq^{s'}  ] = i \delta(\nk - \nq) \delta_{s s'}
\end{equation}
where $s=$R,I and $\delta_{s s'}$ is Kronecker's delta.

The quantum ``free'' Hamiltonian term corresponding to \eqref{csltermino1b} is then
\begin{equation}\label{csltermino1RI}
	\int_{\mathbb{R}^{3+}} d^3 k  \: a^4  \hat \mH_\nk =  	 \int_{\mathbb{R}^{3+}} d^3 k (\hat H^\text{R}_\nk  + \hat H^\text{I}_\nk )
\end{equation}
with the following definitions
\begin{equation}\label{HamiltRI}
	\hat H^{R,I}_\mathbf{k } \equiv \frac{(\hat p_\mathbf{k }^{R,I} )^2  }{2 }+ \frac{(\hat y_\mathbf{k }^{R,I} )^2  }{2 }  \left( k^2 - \frac{a''}{a}  + a^2  V_{, \phi \phi}    \right).
\end{equation}

In the Schr\"odinger approach, the quantum state of the system is described by a wave functional, $\Phi[y(\x,\eta)]$. In Fourier space (and since the theory is still free in the sense that it does not contain terms with power higher than two in the Lagrangian), the wave functional can also be factorized into mode components as
\begin{equation}\label{wavefunctionmodos}
	\Phi[y(\x,\eta)] = \prod_{\nk}  \Phi_\nk ( y_\nk^\text{R} , y_\nk^\text{I} ) =  \prod_{\nk}  \Phi_\nk^\text{R} ( y_\nk^\text{R} )  \Phi_\nk^\text{I} ( y_\nk^\text{I} ).
\end{equation}
In the field representation, the operators would take the form:
\begin{equation}
	\hat y_\nk^\RI \Phi_\nk^\RI =  y_\nk^\RI \Phi_\nk^\RI, \qquad \hat p_\nk^\RI \Phi_\nk^\RI = -i \frac{\partial \Phi_\nk^\RI }{\partial y_\nk^\RI}.
\end{equation}

The decomposition shown in Eq. \eqref{wavefunctionmodos} and the linear evolution equation \eqref{cslmaster} imply that the quantum state of each mode evolves independently (see also \cite{Martin21}). Thus, in Fourier space, the CSL evolution equation of the quantum state, corresponding to the wave functional(s) of each mode $\Phi_\nk^\text{\RI} ( y_\nk^\text{\RI} ) $, is
\begin{eqnarray}\label{cslmasterF}
	|\Phi_{\nk}^{\textrm{R,I}}, \eta \ket &=&  \hat T \exp \bigg\{   \int_\tau^\eta d \eta  \bigg[  -i \hat H^\text{R,I}_\nk  \nn
	&-&  \frac{ a^4 }{4 \lambda } \bigg(    W_{\nk}(\eta)^\RI- 2 \lambda \hat C_{\nk}^\RI    (\eta)   \bigg)^2  \bigg]  \bigg\}  \nn
	&\times&  	|\Phi_{\nk}^{\RI}, \tau \ket
\end{eqnarray}
where the collapse generating operator (in Fourier space) is given as
\begin{equation}\label{colop3}
	\hat C_\mathbf{k }^{\RI}=  e^{-\frac{k^2 }{2 a^2 H^2} }  \left( \frac{2 \epsilon}{ H } \right)^{1/2}  \left(  \frac{\hat p_\mathbf{k }^{\RI}}{a^2} + 5 H \frac{\hat y_\mathbf{k }^{\RI}}{a}       \right)
\end{equation}
and the Probability Rule is
\begin{equation}\label{cslprobabF}
	P(W_{\nk}^{\RI}) dW_{\nk}^{\RI} =   \bra \Phi_{\nk}^{\RI} , \eta | \Phi_{\nk}^{\RI}, \eta \ket \prod_{\eta'=\tau}^{\eta-d\eta} \frac{a^2 d W_{\nk}(\eta')^{\RI}}{\sqrt{2 \pi \lambda/d\eta}}
\end{equation}

We will adopt the standard assumption regarding the vacuum state. That is, at an early conformal time $\tau$, the modes are in their adiabatic ground state, which is a Gaussian centered at zero with certain spread. This ground state is commonly referred to as the Bunch-Davies (BD) vacuum. Taking into account that the initial quantum state is  Gaussian, and  the CSL evolution equation \eqref{cslmasterF} is quadratic in the canonical variables $\hat y_\nk^\text{R,I},  \hat p_\nk^\text{R,I}$, the form of the state vector in the field basis at any time is
\begin{equation}\label{psionday}
	\Phi^{R,I}(\eta,y_{\nk}^{R,I}) = \exp[- A_{k}(\eta)(y_{\nk}^{R,I})^2 +
	B_{k}(\eta)y_{\nk}^{R,I} +  C_{k}(\eta)].
\end{equation}

Therefore, the wave functional evolves according to the CSL equation \eqref{cslmasterF}, with initial conditions given by
\begin{equation}
	A_k (\tau ) = \frac{k}{2}, \qquad B_k (\tau ) = C_k(\tau )= 0.
\end{equation}

Those initial conditions correspond to the BD vacuum, which is perfectly homogeneous and isotropic in the sense of a vacuum state in quantum theory.

Considering that our main goal is to obtain the primordial power spectrum, see Eq. \eqref{masterPS}, we focus our efforts in computing  the quantity $\overline{\bra \hat y_\nk \ket \bra \hat y_\nq \ket^* }$. The expectation values, of course, will be evaluated at the evolved state provided by \eqref{cslmasterF}. In terms of the real and imaginary parts we have
\begin{equation}\label{masterpromedios}
	\overline{\bra  \hat y_\nk \ket \bra \hat y_\nq \ket^* } =  \left( \overline{ \bra \hat  y_\nk^\text{R} \ket^2  }  + \overline{ \bra \hat  y_\nk^\text{I} \ket^2  }         \right) \delta(\nk-\nq ).
\end{equation}
Also, from \eqref{masterpromedios}, it is clear that we are interested in computing the quantities $\overline{ \bra \hat  y_\nk^\text{R,I} \ket^2  } $. In fact, the calculation of the real and imaginary part are exactly the same, so we will only focus on one of them. Additionally, we will simplify the notation by omitting the indexes R,I from now on.

Using the Gaussian wave function \eqref{psionday}, and the two main CSL equations for the mode's quantum state [Eqs. \eqref{cslmasterF} and \eqref{cslprobabF}], we obtain
\begin{equation}\label{masterresta}
	\overline{ \bra \hat  y_\nk \ket^2  } = \overline{ \bra \hat  y_\nk^2 \ket  } - \frac{1}{4 \textrm{Re}[A_k (\eta)]  }.
\end{equation}

\subsection{First term of \eqref{masterresta}}

We now turn our attention to the first term on the right hand side of \eqref{masterresta}, i.e $ \overline{ \bra \hat  y_\nk^2 \ket  }$. The two main CSL equations for the quantum state of each mode,  Eqs. \eqref{cslmasterF} and \eqref{cslprobabF}, can be used to obtain the equation of evolution for the density matrix. This yields one equation per Fourier mode, which can be written as
\begin{equation}\label{densitymatrixevolF}
	\frac{ \partial \hat \rho_\nk (\eta) }{\partial \eta} = - i [\hat H_\nk, \hat \rho_\nk (\eta)] - \frac{\lambda a^4}{2} [\hat C_\nk(\eta), [\hat C_\nk(\eta), \hat \rho_\nk(\eta) ]   ]
\end{equation}
The latter equation can also be obtained form the equation of evolution of the density matrix \eqref{densitymatrixevol} in Fourier space. Thus, it serves as a consistency check.

The importance of Eq. \eqref{densitymatrixevolF} is that one can use it to derive the equation of evolution for the stochastic mean of the quantum expectation value of any operator $\hat O_\nk$. This is,
\begin{equation}\label{vesperadoO}
	\frac{\partial}{\partial \eta} \overline{\bra \hat O_\nk \ket} = -i  \overline{\bra [ \hat O_\nk, \hat H_\nk    ] \ket }  - \frac{a^4 \lambda}{2}   \overline{\bra [ \hat C_\nk , [ \hat C_\nk, \hat O_\nk ]] \ket}
\end{equation}

We take into account that $\lambda$ during inflation is given by Eq. \eqref{lambda2}. This is,
\begin{equation}\label{lambda3}
	\lambda(\eta) = \lambda_0 \left( \frac{a(\eta) }{a_\tau} \right)^{2\alpha}
\end{equation}
Moreover, by choosing $\alpha = 3/2$ and estimating $a_\tau \simeq 1/(H |\tau|) $, we obtain
\begin{equation}\label{key}
	\lambda(\eta) = \lambda_0 H^3 |\tau|^3  a(\eta)^3
\end{equation}

We define the quantities $Q \equiv \overline{ \bra \hat  y_\nk^2 \ket  } $, $R \equiv \overline{ \bra \hat  p_\nk^2 \ket  } $ and $T \equiv \overline{ \bra \hat  p_\nk \hat  y_\nk  + \hat  y_\nk \hat  p_\nk \ket } $, \textit{notation warning:} do not confuse this $R$ with the Ricci scalar. The evolution equations of $Q$, $R$ and $T$ obtained from \eqref{vesperadoO} are:
\begin{subequations}\label{evolQRS}
	\begin{equation}\label{Qprima}
		Q'= T + a^3 \lambda_1 e^{-\frac{k^2 }{ a^2 H^2} }
	\end{equation}
	\begin{equation}\label{Rprima}
		R' = -w_k(\eta)T  + a^5 \lambda_1 25 H^2 e^{-\frac{k^2 }{ a^2 H^2} }
	\end{equation}
	\begin{equation}\label{Sprima}
		T' = 2 R - 2 Q w_k(\eta)
	\end{equation}
\end{subequations}
where
\begin{equation}\label{defweta}
	w_k(\eta) \equiv    k^2 - \frac{a''}{a}  + a^2  V_{, \phi \phi}
\end{equation}
and
\begin{equation}\label{deflambda1}
	\lambda_1 \equiv  \lambda_0 2    \epsilon H^2 |\tau|^3
\end{equation}

We combine Eqs. \eqref{evolQRS} to obtain a single differential equation for $Q$,
\begin{equation}\label{evolQ}
	Q''' + 4 w_k(\eta)  + 2 Q w_k'(\eta) = S(\eta)
\end{equation}
where the source function $S$ is
\begin{eqnarray}
	S&\equiv& 2  \lambda_1 e^{-\frac{k^2 }{ a^2 H^2} }  \bigg[ a^3  \bigg (5 k^2 +w_k  \bigg) \nn
	&+& 31 a^5 H^2  + \frac{2 a   k^4}{H^2} \bigg]
\end{eqnarray}
In Ref. \cite{MartinShadow} it is shown that Eq. \eqref{evolQ} can be solved by introducing the Green function of the free theory,
\begin{equation}\label{defG}
	G (\eta,\teta ) = \frac{1}{W} \left[ y_k^* (\teta)  y_k(\eta)  - y_k(\teta) y_k^* (\eta)    \right] \Theta (\eta-\teta)
\end{equation}
where $y_k(\eta)$ is a solution of
\begin{equation}\label{evoly}
	y''_k(\eta) + \left[ k^2 - w_k (\eta) \right]  y_k(\eta) = 0
\end{equation}
and $W= y_k'  y_k^*  - y_k^{*'} y_k  $ is the Wronskian, which for Eq. \eqref{evoly}  is a constant.  The solution to Eq. \eqref{evolQ} is then
\begin{equation}\label{solQ}
	Q_k (\eta) = |y_k(\eta)|^2 + \frac{1}{2}\int_{\tau}^\eta d\tilde \eta \: S(\tilde \eta) G(\eta,\tilde \eta)^2
\end{equation}
where $\tau \to -\infty$ is the conformal time at the beginning of inflation.

Considering the approximated form of the scale factor during inflation $a(\eta) \simeq -1/H \eta$, with $H \simeq$ constant; one has $w_k(\eta) \simeq k^2 - 2/\eta^2$. The solution of \eqref{evoly} is then
\begin{equation}\label{soly}
	y_k(\eta) = \frac{1}{\sqrt{2k}} \left(     1 + \frac{i}{k \eta}     \right) e^{i k \eta}
\end{equation}
Consequently, the Green function \eqref{defG} is
\begin{eqnarray}\label{G}
	G (\eta,\teta) &=&   \frac{1}{k^3 \eta \teta}  \bigg(     (1+k^2 \eta \teta) \sin [ k (\eta - \teta)]  \nn
	&-& k (\eta-\teta) \cos[(\eta-\teta)]       \bigg) \Theta (\eta - \teta)
\end{eqnarray}
Moreover, in the super-Hubble limit, we have
\begin{equation}\label{Gfh}
	G(\eta,\teta)|_{(-k\eta , -k\teta)\to (0,0)} \simeq \frac{\eta^3 - \teta^3}{3 \eta \teta}  \Theta (\eta - \teta)
\end{equation}

On the other hand, the amplitude of the source function $S(\teta)$ is exponentially suppressed in the regime $k \tau \to -\infty$, while it reaches its maximum in the opposite regime $-k \teta \to 0$. Therefore, when computing the integral in \eqref{solQ}, we can consider the approximated expression \eqref{Gfh};  i.e. we use the form of $G(\eta,\teta)$ in the super-Hubble limit.

Furthermore, we are interested in computing the power spectrum that is used to compare the theoretical predictions with the observational data. Specifically, we have to evaluate $Q$ in the regime of observational interest, i.e. in the super-Hubble limit. After taking this limit, at the leading order we finally obtain
\begin{eqnarray}\label{Qfh2}
	Q_k(\eta) |_{-k\eta \to 0 } &=&  \frac{1}{2 k^3 \eta^2} \bigg\{   1 - \frac{2 \sigma_k}{9} \bigg[ 21 + 29 \gamma_\varepsilon -58 N \nn
	&+& 58 \ln (k |\tau|)  \bigg] \bigg\}
\end{eqnarray}
where $N$ is the total number of e-foldings from the beginning of inflation,  $\gamma_\varepsilon$ is the Euler-Mascheroni constant and the coefficient $\sigma_k$ is defined in \eqref{defsigmak}.

\subsection{Second term of \eqref{masterresta}}

Continuing with the second half of the calculations, we focus now on the second term on the right hand side of \eqref{masterresta}; i.e. the term $ [4 \text{Re}(A_k)  ]^{-1}$. We take the time derivative of \eqref{cslmasterF}, obtaining
\begin{equation}\label{Htotal}
	\frac{\partial}{\partial \eta}|\Phi_{\nk}^{\textrm{R,I}}, \eta \ket = -    i \hat{H}_{\nk}^{\textrm{R,I}}
	+ \hat{H}_{\nk\: \textrm{CSL}}^{\textrm{R,I}}  |\Phi_{\nk}^{\textrm{R,I}}, \eta \ket
\end{equation}
with
\begin{eqnarray}\label{HCSL}
	\hat{H}_{\nk\: \textrm{CSL}}^{\textrm{R,I}} &\equiv& - \frac{a^4 (W^\RI (\nk,\eta))^2}{4 \lambda } + a^4 {W}^\RI (\nk,\eta)  \hat{C}_{\nk}^{\textrm{R,I}} \nn
	&-&  a^4 (\hat{C}_{\nk}^{\textrm{R,I}})^2 \lambda
\end{eqnarray}

We apply the CSL evolution operator, as characterized by Eq. \eqref{Htotal}, to the wave function \eqref{psionday} and regroup terms of order $y^2$, $y^1$ and $y^0$; the evolution equations corresponding to these terms are thus decoupled. Fortunately, the evolution equation corresponding to $y^2$ only contains  $A_k(\eta)$, which is the function we are interested in. The evolution equation is then
\begin{eqnarray}\label{evolAk}
	A_k'  &=& \frac{i w_k }{2}   - 2i A_k^2  +  e^{-\frac{k^2}{ a^2 H^2}} \lambda_1    \bigg( -4 A_k^2    a^3 \nn
	&+& 25 H^2 a^5 + i 20 H A_k a^4 \bigg)
\end{eqnarray}
where $w_k$ and $\lambda_1$  are defined in Eqs. \eqref{defweta} and \eqref{deflambda1} respectively.

Performing the following change of variable
\begin{equation}\label{deff}
	A_k = \frac{1}{N} \left(   \frac{f_k'}{f_k}   -  g \right)
\end{equation}
with
\begin{equation}\label{defN}
	N\equiv 2 (  i + 2  \lambda_1 a^3  e^{-\frac{k^2}{ a^2 H^2}})
\end{equation}
and
\begin{equation}\label{defgk}
	g \equiv  -\frac{1}{2}  \left( \frac{ N'}{N }  + i 20 H \lambda_1 a^4   e^{-\frac{k^2}{ a^2 H^2}} \right)
\end{equation}
equation \eqref{evolAk} can be recast in a more useful way:
\begin{equation}\label{evolf}
	f_k'' + P(\eta) f_k =0
\end{equation}
where
\begin{eqnarray}\label{defPeta}
	P(\eta) &\equiv&  -\frac{i w_k N }{2}  - \left(   g' - g \frac{N'}{N}   \right)  + g^2\nn
	&-&  25H^2 \lambda_1  a^5 N e^{-\frac{k^2}{ a^2 H^2}}    \nn
	&+& i 20 H \lambda_1 a^4 e^{-\frac{k^2}{ a^2 H^2}}  g.
\end{eqnarray}
We note that if $\lambda_1 = 0$ (i.e. no collapse) $f_k$ satisfies the usual equation $f_k'' = -w_k f_k$. The exact solution of  Eq. \eqref{evolf} is difficult to find,  but it can be done perturbatively in $\lambda_1$.  Therefore, at leading order in $\lambda_1$,  the function $P(\eta)$ is explicitly given as
\begin{equation}\label{defPeta2}
	P(\eta) = w_k+ \Sigma(\eta) \lambda_1 + \mathcal{O} (\lambda_1^2)
\end{equation}
where
\begin{eqnarray}\label{defSigma}
	\Sigma(\eta) &\equiv&- \frac{2 i a  }{H^2} 	e^{-\frac{k^2}{ a^2 H^2}}  \bigg( 11 a^4 H^4 \nn
	&-& 5 a^2 H^2 k^2 + 2k^4 + a^2 H^2 w_k \bigg).
\end{eqnarray}

The perturbed solution to Eq. \eqref{evolf} can be  written as:
\begin{equation}\label{solf}
	f_k = f^{(0)}_k + \lambda_1 f^{(1)}_k + \mathcal{O} (\lambda_1^2)
\end{equation}
The function $f^{(0)}_k$ is the solution of the mode equation for $\lambda_1=0$. Specifically, we can choose the normalized mode function in the Bunch-Davies vacuum,  which at leading order in slow roll is given by
\begin{equation}\label{solf0}
	f^{(0)}_k = \frac{1}{\sqrt{2 k }}   \left( 1 + \frac{i}{k \eta}\right) e^{i k \eta}
\end{equation}
Inserting expansions \eqref{defPeta2}, \eqref{solf} in \eqref{evolf}, one finds that the function $f^{(1)}_k$ satisfies
\begin{equation}\label{evolf1}
	f^{(1)''}_k + w_k f^{(1)}_k = -\Sigma(\eta) f^{(0)}_k
\end{equation}

In order to solve Eq. \eqref{evolf1}, we will work at the leading order in slow roll, this is, $a(\eta) \simeq -1/H \eta$, which implies $w_k = k^2-2/\eta^2$. Therefore, the function $\Sigma(\eta)$ is explicitly
\begin{equation}\label{Sigmaeta}
	\Sigma (\eta) = \frac{2 i  	e^{-k^2 \eta^2 }}{\eta^5 H^3} \left(   9 - 4 k^2 \eta^2 + 2 \eta^4 k^4  \right)
\end{equation}

The solution to Eq. \eqref{evolf1} is
\begin{equation}\label{solf10}
	f^{(1)}_k (\eta) = -\int_\tau^\eta d \teta \: \Sigma(\teta)  f^{(0)}_k (\eta) G(\eta,\teta)
\end{equation}
where $G(\eta,\teta)$ is the Green function introduced previously, Eq. \eqref{G}.  As before, the amplitude of the source term is exponentially suppressed in the regime $k \tau \to -\infty$, so the mayor contribution to the integral in Eq. \eqref{solf10} comes from the super-Hubble limit. This means that we can use the approximated expressions for $f^{(0)}_k (\eta)$ and $G(\eta,\teta)$, Eq.\eqref{Gfh}  in such a limit. The result is
\begin{eqnarray}\label{solf1}
	f^{(1)}(x) & =& \frac{k^{5/2}}{H^3 6 \sqrt{2}} \frac{e^{-x^2}}{x^4} \bigg[ 6 - 33x^2 - 11 x^4 \nn
	&+& 11 e^{x^2} x^3 \bigg(     4 \sqrt{\pi } (1- \text{Erf}[x] ) - x^3 \text{Ei}[-x^2] \bigg) \bigg], \nn
\end{eqnarray}
where  we have used the definition $x \equiv - k \eta$. The function Ei$(z)$ is the exponential integral function defined as
\begin{equation}\label{defEi}
	\text{Ei} (z) \equiv - \int_{-z}^\infty dt \: \frac{e^{-t}}{t}
\end{equation}
and Erf$(z)$ is the Gaussian error function
\begin{equation}\label{defErf}
	\text{Erf}(z) \equiv  \frac{2}{\sqrt{\pi}}  \int_0^z dt\:  e^{-t^2}
\end{equation}

Returning to the original variable $A_k$, Eq. \eqref{deff},  with the perturbed solution in terms of Eqs. \eqref{solf0} and \eqref{solf1}, we obtain
\begin{eqnarray}\label{Aksol}
	A_k &=&   \frac{1}{2} k \left(1-\frac{1}{x (x+i)}\right)  +  \frac{\lambda_1   k^4 e^{-x(i+x)} }{12 H^3 x^4 (x+i)^2}  \nn
	&\times&\bigg\{ -18-6 e^{i x} (1+x) (-5i -5x +2x^3)   \nn
	&+&   x \left[   18 i + x (15-21ix +11 x^3 (3i+x) )  \right] + 11 e^{x^2} x^5  \nn
	&\times&  \left(   -4 \sqrt{\pi} \text{Erf}[x] + x (-3 + x (3i + x))  \text{Ei}[-x^2]    \right)  \biggr\}  \nn
	&+& \mathcal{O} (\lambda_1^2)
\end{eqnarray}

From the observational point of view, we are interested in taking the limit $ x \to 0$. Retaining the dominant terms in the perturbed solutions, we find that the real part of $A_k$ can be approximated by
\begin{equation}\label{ReAk}
	\text{Re}(A_k)|_{x \to 0}  \simeq \frac{k x^2}{2} \left(  1 + 8 \frac{\lambda_1 k^3}{H^3 x^6}    \right)
\end{equation}
In fact, we rewrite the previous expression using the coefficient $\sigma_k$ defined in Eq. \eqref{defsigmak}, which yields
\begin{equation}\label{ReAk2}
	\text{Re}(A_k)|_{x \to 0}  \simeq \frac{k x^2}{2} \left(  1 + 16  \frac{\sigma_k}{x^6}    \right)
\end{equation}
We also introduce the function
\begin{equation}\label{defF}
	F(\sigma_k, x)|_{x \to 0} \equiv	\frac{2 k x^2}{4 \text{Re}(A_k)|_{x \to 0}}. \:
\end{equation}
As we can see from Eq. \eqref{ReAk2},  in the limit  $x\to0$,  the function  $F(\sigma_k, x)|_{x \to 0} $ goes to zero as $x^6$, $\:$ i.e.
$$|F(\sigma_k, x)|_{x \to 0}  \ll 1$$
On the other hand, if there is no collapse, this is, if $\lambda=0$,  then $\sigma_k = 0$ and $F(0, x)|_{x \to 0} = 1$

It is now straightforward to write the explicit expression for the equivalent power spectrum. Substituting Eqs. \eqref{Qfh2} and \eqref{defF} in \eqref{masterresta}, and making use of Eq. \eqref{masterPS},  we obtain the primordial power spectrum as shown in Eq. \eqref{PS0}.

\section{Dynamical evolution of subhorizon (and superhorizon) modes}\label{AppB}

In this Appendix, we provide an analysis regarding the evolution of the subhorizon modes as dictated by the CSL evolution equation. For ease of the calculations,  we will rely on the evolution equation for the density matrix Eq. \eqref{densitymatrixevolF},  the analysis of course would be equivalent if the CSL master equation \eqref{cslmasterF} was used instead.  Therefore, from Eq. \eqref{densitymatrixevolF}, we observe that the term affecting the dynamical evolution of any mode, due to the CSL process, is
\begin{equation}\label{defD}
	\hat D^{\text{CSL}}_\nk \equiv \frac{\lambda(a) a^4}{2} [\hat C_\nk(a), [\hat C_\nk(a), \hat \rho_\nk ]   ].
\end{equation}
Additionally, the collapse rate parameter $\lambda$ as given in Eq. \eqref{lambda2} can be recast as
\begin{equation}\label{lambda2a}
	\lambda(a) = \lambda_0 \left( \frac{a}{a_\tau}\right)^{3},
\end{equation}
where in the above equation we have used  $\alpha = 3/2$ (we have chosen this value due to the results of \ref{AppA}) and the definition  of $N$, i.e. the number of e-folds since the beginning of inflation.

Substituting Eq. \eqref{lambda2a} and the expression for the collapse operator $\hat{C}_{\nk}$, as given in Eq. \eqref{colopfourier}, we can rewrite $\hat D^{\text{CSL}}_\nk $ as

\begin{eqnarray}\label{D2}
	\hat D^{\text{CSL}}_\nk  &=& \lambda_0 \frac{\epsilon}{H} \left(  \frac{a}{a_\tau}\right)^3 e^{-k^2/a^2 H^2} \bigg\{     [ \hat p_{\nk} ,[ \hat p_{\nk},  \hat \rho_\nk]]    \nn
	&+&  5 H a \bigg(   [ \hat y_{\nk} ,[ \hat p_{\nk},  \hat \rho_\nk]]  + [ \hat p_{\nk} ,[ \hat y_{\nk},  \hat \rho_\nk]]        \bigg)  \nn
	&+& 25 H^2 a^2 [ \hat y_{\nk} ,[ \hat y_{\nk},  \hat \rho_\nk]]    \bigg\}.
\end{eqnarray}

We proceed to analyze  the last term in the previous equation, i.e. the one with the highest power of $a$. The analysis corresponding to the other terms is equivalent. Thus, the dynamical evolution for the last term is dictated by the function
\begin{equation}\label{defgamma}
\Delta_k(a) \equiv a^5 e^{-k^2/a^2 H^2}.
\end{equation}
We note that $\Delta_k (a) < 1$ if and only if $\kappa^2 > 5 \ln a$, where $\kappa \equiv k/a H$.  Assuming that inflation lasts approximately 65 e-folds, then $a \sim e^{65}$. As a consequence,  $\Delta_k (a) < 1$ if and only if $\kappa > \mathcal{O}(10)$. In particular, for a subhorizon mode $\kappa \gg 1$, one has that $\Delta_k (a) \ll 1$.  This demonstrates that, for subhorizon modes,  the additional dynamical evolution introduced by the CSL mechanism is exponentially suppressed.

On the other hand, an analogous analysis for superhorizon modes $\kappa \ll 1$, indicates that $\Delta_k (a) \gg 1$, i.e. the superhorizon modes are the most affected by the CSL evolution term.


\bibliography{bibliografia}   

\begin{thebibliography}{115}
\expandafter\ifx\csname natexlab\endcsname\relax\def\natexlab#1{#1}\fi
\expandafter\ifx\csname bibnamefont\endcsname\relax
  \def\bibnamefont#1{#1}\fi
\expandafter\ifx\csname bibfnamefont\endcsname\relax
  \def\bibfnamefont#1{#1}\fi
\expandafter\ifx\csname citenamefont\endcsname\relax
  \def\citenamefont#1{#1}\fi
\expandafter\ifx\csname url\endcsname\relax
  \def\url#1{\texttt{#1}}\fi
\expandafter\ifx\csname urlprefix\endcsname\relax\def\urlprefix{URL }\fi
\providecommand{\bibinfo}[2]{#2}
\providecommand{\eprint}[2][]{\url{#2}}

\bibitem[{\citenamefont{Starobinsky}(1979)}]{Starobinsky79}
\bibinfo{author}{\bibfnamefont{A.~A.} \bibnamefont{Starobinsky}},
  \bibinfo{journal}{JETP Lett.} \textbf{\bibinfo{volume}{30}},
  \bibinfo{pages}{682} (\bibinfo{year}{1979}).

\bibitem[{\citenamefont{Starobinsky}(1980)}]{Starobinsky80}
\bibinfo{author}{\bibfnamefont{A.~A.} \bibnamefont{Starobinsky}},
  \bibinfo{journal}{Phys. Lett.} \textbf{\bibinfo{volume}{B91}},
  \bibinfo{pages}{99} (\bibinfo{year}{1980}).

\bibitem[{\citenamefont{Guth}(1981)}]{Guth81}
\bibinfo{author}{\bibfnamefont{A.~H.} \bibnamefont{Guth}},
  \bibinfo{journal}{Phys. Rev.} \textbf{\bibinfo{volume}{D23}},
  \bibinfo{pages}{347} (\bibinfo{year}{1981}).

\bibitem[{\citenamefont{Mukhanov and Chibisov}(1981)}]{Mukhanov81}
\bibinfo{author}{\bibfnamefont{V.~F.} \bibnamefont{Mukhanov}} \bibnamefont{and}
  \bibinfo{author}{\bibfnamefont{G.~V.} \bibnamefont{Chibisov}},
  \bibinfo{journal}{JETP Lett.} \textbf{\bibinfo{volume}{33}},
  \bibinfo{pages}{532} (\bibinfo{year}{1981}), \bibinfo{note}{[Pisma Zh. Eksp.
  Teor. Fiz.33,549(1981)]}.

\bibitem[{\citenamefont{Linde}(1982)}]{Linde82}
\bibinfo{author}{\bibfnamefont{A.~D.} \bibnamefont{Linde}},
  \bibinfo{journal}{Phys. Lett.} \textbf{\bibinfo{volume}{B108}},
  \bibinfo{pages}{389} (\bibinfo{year}{1982}).

\bibitem[{\citenamefont{Linde}(1983)}]{Linde83}
\bibinfo{author}{\bibfnamefont{A.~D.} \bibnamefont{Linde}},
  \bibinfo{journal}{Phys. Lett. B} \textbf{\bibinfo{volume}{129}},
  \bibinfo{pages}{177} (\bibinfo{year}{1983}).

\bibitem[{\citenamefont{Albrecht and Steinhardt}(1982)}]{Albrecht82}
\bibinfo{author}{\bibfnamefont{A.}~\bibnamefont{Albrecht}} \bibnamefont{and}
  \bibinfo{author}{\bibfnamefont{P.~J.} \bibnamefont{Steinhardt}},
  \bibinfo{journal}{Phys. Rev. Lett.} \textbf{\bibinfo{volume}{48}},
  \bibinfo{pages}{1220} (\bibinfo{year}{1982}).

\bibitem[{\citenamefont{Bardeen et~al.}(1983)\citenamefont{Bardeen, Steinhardt,
  and Turner}}]{Bardeen83}
\bibinfo{author}{\bibfnamefont{J.~M.} \bibnamefont{Bardeen}},
  \bibinfo{author}{\bibfnamefont{P.~J.} \bibnamefont{Steinhardt}},
  \bibnamefont{and} \bibinfo{author}{\bibfnamefont{M.~S.}
  \bibnamefont{Turner}}, \bibinfo{journal}{Phys. Rev.}
  \textbf{\bibinfo{volume}{D28}}, \bibinfo{pages}{679} (\bibinfo{year}{1983}).

\bibitem[{\citenamefont{{Brandenberger}}(1984)}]{Brandenberger84}
\bibinfo{author}{\bibfnamefont{R.~H.} \bibnamefont{{Brandenberger}}},
  \bibinfo{journal}{Nuclear Physics B} \textbf{\bibinfo{volume}{245}},
  \bibinfo{pages}{328} (\bibinfo{year}{1984}).

\bibitem[{\citenamefont{Hawking}(1982)}]{Hawking82}
\bibinfo{author}{\bibfnamefont{S.~W.} \bibnamefont{Hawking}},
  \bibinfo{journal}{Phys. Lett.} \textbf{\bibinfo{volume}{115B}},
  \bibinfo{pages}{295} (\bibinfo{year}{1982}).

\bibitem[{\citenamefont{Aghanim et~al.}(2020{\natexlab{a}})}]{Planck18a}
\bibinfo{author}{\bibfnamefont{N.}~\bibnamefont{Aghanim}} \bibnamefont{et~al.}
  (\bibinfo{collaboration}{Planck}), \bibinfo{journal}{Astron. Astrophys.}
  \textbf{\bibinfo{volume}{641}}, \bibinfo{pages}{A1}
  (\bibinfo{year}{2020}{\natexlab{a}}), \eprint{1807.06205}.

\bibitem[{\citenamefont{Akrami et~al.}(2020)}]{Planck18b}
\bibinfo{author}{\bibfnamefont{Y.}~\bibnamefont{Akrami}} \bibnamefont{et~al.}
  (\bibinfo{collaboration}{Planck}), \bibinfo{journal}{Astron. Astrophys.}
  \textbf{\bibinfo{volume}{641}}, \bibinfo{pages}{A10} (\bibinfo{year}{2020}),
  \eprint{1807.06211}.

\bibitem[{\citenamefont{Aghanim et~al.}(2020{\natexlab{b}})}]{Planck18c}
\bibinfo{author}{\bibfnamefont{N.}~\bibnamefont{Aghanim}} \bibnamefont{et~al.}
  (\bibinfo{collaboration}{Planck}), \bibinfo{journal}{Astron. Astrophys.}
  \textbf{\bibinfo{volume}{641}}, \bibinfo{pages}{A6}
  (\bibinfo{year}{2020}{\natexlab{b}}), \eprint{1807.06209}.

\bibitem[{\citenamefont{{Wigner}}(1963)}]{Wigner63}
\bibinfo{author}{\bibfnamefont{E.~P.} \bibnamefont{{Wigner}}},
  \bibinfo{journal}{American Journal of Physics} \textbf{\bibinfo{volume}{31}},
  \bibinfo{pages}{6} (\bibinfo{year}{1963}).

\bibitem[{\citenamefont{Omnes}(1994)}]{Omnes}
\bibinfo{author}{\bibfnamefont{R.}~\bibnamefont{Omnes}},
  \emph{\bibinfo{title}{The Interpretation of Quantum Mechanics}}
  (\bibinfo{publisher}{Princeton University Press}, \bibinfo{year}{1994}).

\bibitem[{\citenamefont{Albert}(1994)}]{Albert}
\bibinfo{author}{\bibfnamefont{D.~Z.} \bibnamefont{Albert}},
  \bibinfo{journal}{\emph{Quantum Mechanics and Experience}, (Harvard
  University Press)}  (\bibinfo{year}{1994}).

\bibitem[{\citenamefont{Maudlin}(1995)}]{Maudlin95}
\bibinfo{author}{\bibfnamefont{T.}~\bibnamefont{Maudlin}},
  \bibinfo{journal}{Topoi} \textbf{\bibinfo{volume}{14}}, \bibinfo{pages}{7}
  (\bibinfo{year}{1995}).

\bibitem[{\citenamefont{Becker}(2018)}]{Becker}
\bibinfo{author}{\bibfnamefont{A.}~\bibnamefont{Becker}},
  \emph{\bibinfo{title}{What is Real? The Unfinished Quest for the Meaning of
  Quantum Physics}} (\bibinfo{publisher}{Basic Books, New York},
  \bibinfo{year}{2018}).

\bibitem[{\citenamefont{Norsen}(2017)}]{Norsen}
\bibinfo{author}{\bibfnamefont{T.}~\bibnamefont{Norsen}},
  \emph{\bibinfo{title}{{Foundations of Quantum Mechanics}}}
  (\bibinfo{publisher}{Springer International Publishing AG},
  \bibinfo{year}{2017}).

\bibitem[{\citenamefont{Durr and Lazarovici}(2020)}]{Durr}
\bibinfo{author}{\bibfnamefont{D.}~\bibnamefont{Durr}} \bibnamefont{and}
  \bibinfo{author}{\bibfnamefont{D.}~\bibnamefont{Lazarovici}},
  \emph{\bibinfo{title}{{Understanding Quantum Mechanics}}}
  (\bibinfo{publisher}{Springer International Publishing AG},
  \bibinfo{year}{2020}).

\bibitem[{\citenamefont{Okon}(2014)}]{okon14}
\bibinfo{author}{\bibfnamefont{E.}~\bibnamefont{Okon}}, \bibinfo{journal}{Rev.
  Mex. Fis. E} \textbf{\bibinfo{volume}{60}}, \bibinfo{pages}{130}
  (\bibinfo{year}{2014}).

\bibitem[{\citenamefont{Bell}(1981)}]{Bell81}
\bibinfo{author}{\bibfnamefont{J.~S.} \bibnamefont{Bell}},
  \emph{\bibinfo{title}{Quantum Mechanics for cosmologists}}, Quantum Gravity 2
  (\bibinfo{publisher}{eds. Isham, C., Penrose, R. and Sciama, D., Oxford
  University Press}, \bibinfo{year}{1981}).

\bibitem[{\citenamefont{Perez et~al.}(2006)\citenamefont{Perez, Sahlmann, and
  Sudarsky}}]{PSS06}
\bibinfo{author}{\bibfnamefont{A.}~\bibnamefont{Perez}},
  \bibinfo{author}{\bibfnamefont{H.}~\bibnamefont{Sahlmann}}, \bibnamefont{and}
  \bibinfo{author}{\bibfnamefont{D.}~\bibnamefont{Sudarsky}},
  \bibinfo{journal}{Class. Quant. Grav.} \textbf{\bibinfo{volume}{23}},
  \bibinfo{pages}{2317} (\bibinfo{year}{2006}), \eprint{gr-qc/0508100}.

\bibitem[{\citenamefont{Sudarsky}(2011)}]{Sudarsky11}
\bibinfo{author}{\bibfnamefont{D.}~\bibnamefont{Sudarsky}},
  \bibinfo{journal}{Int. J. Mod. Phys. D} \textbf{\bibinfo{volume}{20}},
  \bibinfo{pages}{509} (\bibinfo{year}{2011}), \eprint{0906.0315}.

\bibitem[{\citenamefont{Landau et~al.}(2013)\citenamefont{Landau, Le\'on, and
  Sudarsky}}]{Susana13}
\bibinfo{author}{\bibfnamefont{S.}~\bibnamefont{Landau}},
  \bibinfo{author}{\bibfnamefont{G.}~\bibnamefont{Le\'on}}, \bibnamefont{and}
  \bibinfo{author}{\bibfnamefont{D.}~\bibnamefont{Sudarsky}},
  \bibinfo{journal}{Phys. Rev. D} \textbf{\bibinfo{volume}{88}},
  \bibinfo{pages}{023526} (\bibinfo{year}{2013}), \eprint{1107.3054}.

\bibitem[{\citenamefont{Bengochea}(2020)}]{Bengochea20}
\bibinfo{author}{\bibfnamefont{G.~R.} \bibnamefont{Bengochea}},
  \bibinfo{journal}{Rev. Mex. Fis. E} \textbf{\bibinfo{volume}{17}},
  \bibinfo{pages}{263} (\bibinfo{year}{2020}), \eprint{2007.03428}.

\bibitem[{\citenamefont{Berjon et~al.}(2021)\citenamefont{Berjon, Okon, and
  Sudarsky}}]{Berjon21}
\bibinfo{author}{\bibfnamefont{J.}~\bibnamefont{Berjon}},
  \bibinfo{author}{\bibfnamefont{E.}~\bibnamefont{Okon}}, \bibnamefont{and}
  \bibinfo{author}{\bibfnamefont{D.}~\bibnamefont{Sudarsky}},
  \bibinfo{journal}{Phys. Rev. D} \textbf{\bibinfo{volume}{103}},
  \bibinfo{pages}{043521} (\bibinfo{year}{2021}), \eprint{2009.09999}.

\bibitem[{\citenamefont{{Hartle}}(1993)}]{Hartle93}
\bibinfo{author}{\bibfnamefont{J.~B.} \bibnamefont{{Hartle}}},
  \bibinfo{journal}{arXiv e-prints} \bibinfo{eid}{gr-qc/9304006}
  (\bibinfo{year}{1993}), \eprint{gr-qc/9304006}.

\bibitem[{\citenamefont{Kiefer and Polarski}(2009)}]{kiefer09}
\bibinfo{author}{\bibfnamefont{C.}~\bibnamefont{Kiefer}} \bibnamefont{and}
  \bibinfo{author}{\bibfnamefont{D.}~\bibnamefont{Polarski}},
  \bibinfo{journal}{Adv. Sci. Lett.} \textbf{\bibinfo{volume}{2}},
  \bibinfo{pages}{164} (\bibinfo{year}{2009}), \eprint{0810.0087}.

\bibitem[{\citenamefont{Halliwell}(1989)}]{halliwell}
\bibinfo{author}{\bibfnamefont{J.~J.} \bibnamefont{Halliwell}},
  \bibinfo{journal}{Phys. Rev.} \textbf{\bibinfo{volume}{D39}},
  \bibinfo{pages}{2912} (\bibinfo{year}{1989}).

\bibitem[{\citenamefont{Kiefer}(2000)}]{kiefer2}
\bibinfo{author}{\bibfnamefont{C.}~\bibnamefont{Kiefer}},
  \bibinfo{journal}{Nucl. Phys. Proc. Suppl.} \textbf{\bibinfo{volume}{88}},
  \bibinfo{pages}{255} (\bibinfo{year}{2000}), \eprint{astro-ph/0006252}.

\bibitem[{\citenamefont{Polarski and Starobinsky}(1996)}]{polarski}
\bibinfo{author}{\bibfnamefont{D.}~\bibnamefont{Polarski}} \bibnamefont{and}
  \bibinfo{author}{\bibfnamefont{A.~A.} \bibnamefont{Starobinsky}},
  \bibinfo{journal}{Class. Quant. Grav.} \textbf{\bibinfo{volume}{13}},
  \bibinfo{pages}{377} (\bibinfo{year}{1996}), \eprint{gr-qc/9504030}.

\bibitem[{\citenamefont{Okon and Sudarsky}(2016{\natexlab{a}})}]{okon16}
\bibinfo{author}{\bibfnamefont{E.}~\bibnamefont{Okon}} \bibnamefont{and}
  \bibinfo{author}{\bibfnamefont{D.}~\bibnamefont{Sudarsky}},
  \bibinfo{journal}{Found. Phys.} \textbf{\bibinfo{volume}{46}},
  \bibinfo{pages}{852} (\bibinfo{year}{2016}{\natexlab{a}}),
  \eprint{1512.05298}.

\bibitem[{\citenamefont{{Adler}}(2003)}]{Adler01}
\bibinfo{author}{\bibfnamefont{S.~L.} \bibnamefont{{Adler}}},
  \bibinfo{journal}{Studies in History and Philosophy of Modern Physics}
  \textbf{\bibinfo{volume}{34}}, \bibinfo{pages}{135} (\bibinfo{year}{2003}),
  \eprint{quant-ph/0112095}.

\bibitem[{\citenamefont{Schlosshauer}(2004)}]{schlosshauer}
\bibinfo{author}{\bibfnamefont{M.}~\bibnamefont{Schlosshauer}},
  \bibinfo{journal}{Rev. Mod. Phys.} \textbf{\bibinfo{volume}{76}},
  \bibinfo{pages}{1267} (\bibinfo{year}{2004}), \eprint{quant-ph/0312059}.

\bibitem[{\citenamefont{{Kent}}(1990)}]{kent}
\bibinfo{author}{\bibfnamefont{A.}~\bibnamefont{{Kent}}},
  \bibinfo{journal}{International Journal of Modern Physics A}
  \textbf{\bibinfo{volume}{5}}, \bibinfo{pages}{1745} (\bibinfo{year}{1990}),
  \eprint{gr-qc/9703089}.

\bibitem[{\citenamefont{{Stapp}}(2002)}]{stapp}
\bibinfo{author}{\bibfnamefont{H.~P.} \bibnamefont{{Stapp}}},
  \bibinfo{journal}{Canadian Journal of Physics} \textbf{\bibinfo{volume}{80}},
  \bibinfo{pages}{1043} (\bibinfo{year}{2002}), \eprint{quant-ph/0110148}.

\bibitem[{\citenamefont{Bohm}(1952)}]{bohm}
\bibinfo{author}{\bibfnamefont{D.}~\bibnamefont{Bohm}}, \bibinfo{journal}{Phys.
  Rev.} \textbf{\bibinfo{volume}{85}}, \bibinfo{pages}{166}
  (\bibinfo{year}{1952}).

\bibitem[{\citenamefont{Valentini}(2008)}]{Valentini08a}
\bibinfo{author}{\bibfnamefont{A.}~\bibnamefont{Valentini}},
  \bibinfo{journal}{arXiv e-prints} \bibinfo{eid}{arXiv:0804.4656}
  (\bibinfo{year}{2008}), \eprint{0804.4656}.

\bibitem[{\citenamefont{Valentini}(2010)}]{Valentini08b}
\bibinfo{author}{\bibfnamefont{A.}~\bibnamefont{Valentini}},
  \bibinfo{journal}{Phys. Rev. D} \textbf{\bibinfo{volume}{82}},
  \bibinfo{pages}{063513} (\bibinfo{year}{2010}), \eprint{0805.0163}.

\bibitem[{\citenamefont{{Pinto-Neto} et~al.}(2012)\citenamefont{{Pinto-Neto},
  {Santos}, and {Struyve}}}]{Neto12}
\bibinfo{author}{\bibfnamefont{N.}~\bibnamefont{{Pinto-Neto}}},
  \bibinfo{author}{\bibfnamefont{G.}~\bibnamefont{{Santos}}}, \bibnamefont{and}
  \bibinfo{author}{\bibfnamefont{W.}~\bibnamefont{{Struyve}}},
  \bibinfo{journal}{Phys. Rev. D} \textbf{\bibinfo{volume}{85}},
  \bibinfo{eid}{083506} (\bibinfo{year}{2012}), \eprint{1110.1339}.

\bibitem[{\citenamefont{{Goldstein} et~al.}(2015)\citenamefont{{Goldstein},
  {Struyve}, and {Tumulka}}}]{goldstein15}
\bibinfo{author}{\bibfnamefont{S.}~\bibnamefont{{Goldstein}}},
  \bibinfo{author}{\bibfnamefont{W.}~\bibnamefont{{Struyve}}},
  \bibnamefont{and}
  \bibinfo{author}{\bibfnamefont{R.}~\bibnamefont{{Tumulka}}},
  \bibinfo{journal}{arXiv e-prints} \bibinfo{eid}{arXiv:1508.01017}
  (\bibinfo{year}{2015}), \eprint{1508.01017}.

\bibitem[{\citenamefont{{Pinto-Neto} and {Struyve}}(2018)}]{Neto18}
\bibinfo{author}{\bibfnamefont{N.}~\bibnamefont{{Pinto-Neto}}}
  \bibnamefont{and}
  \bibinfo{author}{\bibfnamefont{W.}~\bibnamefont{{Struyve}}},
  \bibinfo{journal}{arXiv e-prints} \bibinfo{eid}{arXiv:1801.03353}
  (\bibinfo{year}{2018}), \eprint{1801.03353}.

\bibitem[{\citenamefont{Vitenti et~al.}(2019)\citenamefont{Vitenti, Peter, and
  Valentini}}]{Valentini19}
\bibinfo{author}{\bibfnamefont{S.~D.~P.} \bibnamefont{Vitenti}},
  \bibinfo{author}{\bibfnamefont{P.}~\bibnamefont{Peter}}, \bibnamefont{and}
  \bibinfo{author}{\bibfnamefont{A.}~\bibnamefont{Valentini}},
  \bibinfo{journal}{Phys. Rev. D} \textbf{\bibinfo{volume}{100}},
  \bibinfo{pages}{043506} (\bibinfo{year}{2019}), \eprint{1901.08885}.

\bibitem[{\citenamefont{Pearle}(1976)}]{Pearle76}
\bibinfo{author}{\bibfnamefont{P.}~\bibnamefont{Pearle}},
  \bibinfo{journal}{Phys. Rev. D} \textbf{\bibinfo{volume}{13}},
  \bibinfo{pages}{857} (\bibinfo{year}{1976}).

\bibitem[{\citenamefont{Ghirardi et~al.}(1986)\citenamefont{Ghirardi, Rimini,
  and Weber}}]{Ghirardi86}
\bibinfo{author}{\bibfnamefont{G.}~\bibnamefont{Ghirardi}},
  \bibinfo{author}{\bibfnamefont{A.}~\bibnamefont{Rimini}}, \bibnamefont{and}
  \bibinfo{author}{\bibfnamefont{T.}~\bibnamefont{Weber}},
  \bibinfo{journal}{Phys.Rev.} \textbf{\bibinfo{volume}{D34}},
  \bibinfo{pages}{470} (\bibinfo{year}{1986}).

\bibitem[{\citenamefont{Pearle}(1989)}]{Pearle89}
\bibinfo{author}{\bibfnamefont{P.~M.} \bibnamefont{Pearle}},
  \bibinfo{journal}{Phys.Rev.} \textbf{\bibinfo{volume}{A39}},
  \bibinfo{pages}{2277} (\bibinfo{year}{1989}).

\bibitem[{\citenamefont{Diosi}(1987)}]{Diosi87}
\bibinfo{author}{\bibfnamefont{L.}~\bibnamefont{Diosi}},
  \bibinfo{journal}{Phys.Lett.} \textbf{\bibinfo{volume}{A120}},
  \bibinfo{pages}{377} (\bibinfo{year}{1987}).

\bibitem[{\citenamefont{Diosi}(1989)}]{Diosi89}
\bibinfo{author}{\bibfnamefont{L.}~\bibnamefont{Diosi}},
  \bibinfo{journal}{Phys.Rev.} \textbf{\bibinfo{volume}{A40}},
  \bibinfo{pages}{1165} (\bibinfo{year}{1989}).

\bibitem[{\citenamefont{Penrose}(1996)}]{Penrose96}
\bibinfo{author}{\bibfnamefont{R.}~\bibnamefont{Penrose}},
  \bibinfo{journal}{Gen.Rel.Grav.} \textbf{\bibinfo{volume}{28}},
  \bibinfo{pages}{581} (\bibinfo{year}{1996}).

\bibitem[{\citenamefont{{Bassi} and {Ghirardi}}(2003)}]{Bassi1}
\bibinfo{author}{\bibfnamefont{A.}~\bibnamefont{{Bassi}}} \bibnamefont{and}
  \bibinfo{author}{\bibfnamefont{G.}~\bibnamefont{{Ghirardi}}},
  \bibinfo{journal}{Phys. Rept.} \textbf{\bibinfo{volume}{379}},
  \bibinfo{pages}{257} (\bibinfo{year}{2003}), \eprint{quant-ph/0302164}.

\bibitem[{\citenamefont{{Bassi} et~al.}(2013)\citenamefont{{Bassi}, {Lochan},
  {Satin}, {Singh}, and {Ulbricht}}}]{Bassi2}
\bibinfo{author}{\bibfnamefont{A.}~\bibnamefont{{Bassi}}},
  \bibinfo{author}{\bibfnamefont{K.}~\bibnamefont{{Lochan}}},
  \bibinfo{author}{\bibfnamefont{S.}~\bibnamefont{{Satin}}},
  \bibinfo{author}{\bibfnamefont{T.~P.} \bibnamefont{{Singh}}},
  \bibnamefont{and}
  \bibinfo{author}{\bibfnamefont{H.}~\bibnamefont{{Ulbricht}}},
  \bibinfo{journal}{Reviews of Modern Physics} \textbf{\bibinfo{volume}{85}},
  \bibinfo{pages}{471} (\bibinfo{year}{2013}), \eprint{1204.4325}.

\bibitem[{\citenamefont{Leon and Sudarsky}(2010)}]{Daniel10}
\bibinfo{author}{\bibfnamefont{G.}~\bibnamefont{Leon}} \bibnamefont{and}
  \bibinfo{author}{\bibfnamefont{D.}~\bibnamefont{Sudarsky}},
  \bibinfo{journal}{Class. Quant. Grav.} \textbf{\bibinfo{volume}{27}},
  \bibinfo{pages}{225017} (\bibinfo{year}{2010}), \eprint{1003.5950}.

\bibitem[{\citenamefont{Diez-Tejedor and Sudarsky}(2012)}]{Tejedor12}
\bibinfo{author}{\bibfnamefont{A.}~\bibnamefont{Diez-Tejedor}}
  \bibnamefont{and} \bibinfo{author}{\bibfnamefont{D.}~\bibnamefont{Sudarsky}},
  \bibinfo{journal}{JCAP} \textbf{\bibinfo{volume}{07}}, \bibinfo{pages}{045}
  (\bibinfo{year}{2012}), \eprint{1108.4928}.

\bibitem[{\citenamefont{{Diez-Tejedor}
  et~al.}(2012)\citenamefont{{Diez-Tejedor}, {Le{\'o}n}, and
  {Sudarsky}}}]{Tejedor12B}
\bibinfo{author}{\bibfnamefont{A.}~\bibnamefont{{Diez-Tejedor}}},
  \bibinfo{author}{\bibfnamefont{G.}~\bibnamefont{{Le{\'o}n}}},
  \bibnamefont{and}
  \bibinfo{author}{\bibfnamefont{D.}~\bibnamefont{{Sudarsky}}},
  \bibinfo{journal}{General Relativity and Gravitation}
  \textbf{\bibinfo{volume}{44}}, \bibinfo{pages}{2965} (\bibinfo{year}{2012}),
  \eprint{1106.1176}.

\bibitem[{\citenamefont{Martin et~al.}(2012)\citenamefont{Martin, Vennin, and
  Peter}}]{Martin12}
\bibinfo{author}{\bibfnamefont{J.}~\bibnamefont{Martin}},
  \bibinfo{author}{\bibfnamefont{V.}~\bibnamefont{Vennin}}, \bibnamefont{and}
  \bibinfo{author}{\bibfnamefont{P.}~\bibnamefont{Peter}},
  \bibinfo{journal}{Phys. Rev. D} \textbf{\bibinfo{volume}{86}},
  \bibinfo{pages}{103524} (\bibinfo{year}{2012}), \eprint{1207.2086}.

\bibitem[{\citenamefont{{Ca{\~n}ate} et~al.}(2013)\citenamefont{{Ca{\~n}ate},
  {Pearle}, and {Sudarsky}}}]{Pedro13}
\bibinfo{author}{\bibfnamefont{P.}~\bibnamefont{{Ca{\~n}ate}}},
  \bibinfo{author}{\bibfnamefont{P.}~\bibnamefont{{Pearle}}}, \bibnamefont{and}
  \bibinfo{author}{\bibfnamefont{D.}~\bibnamefont{{Sudarsky}}},
  \bibinfo{journal}{Phys.Rev. D} \textbf{\bibinfo{volume}{87}},
  \bibinfo{eid}{104024} (\bibinfo{year}{2013}), \eprint{1211.3463}.

\bibitem[{\citenamefont{Das et~al.}(2013)\citenamefont{Das, Lochan, Sahu, and
  Singh}}]{Das13}
\bibinfo{author}{\bibfnamefont{S.}~\bibnamefont{Das}},
  \bibinfo{author}{\bibfnamefont{K.}~\bibnamefont{Lochan}},
  \bibinfo{author}{\bibfnamefont{S.}~\bibnamefont{Sahu}}, \bibnamefont{and}
  \bibinfo{author}{\bibfnamefont{T.~P.} \bibnamefont{Singh}},
  \bibinfo{journal}{Phys. Rev. D} \textbf{\bibinfo{volume}{88}},
  \bibinfo{pages}{085020} (\bibinfo{year}{2013}), \bibinfo{note}{[Erratum:
  Phys.Rev.D 89, 109902 (2014)]}, \eprint{1304.5094}.

\bibitem[{\citenamefont{Bengochea et~al.}(2015)\citenamefont{Bengochea,
  {Ca{\~n}ate}, and Sudarsky}}]{Bengochea15}
\bibinfo{author}{\bibfnamefont{G.~R.} \bibnamefont{Bengochea}},
  \bibinfo{author}{\bibfnamefont{P.}~\bibnamefont{{Ca{\~n}ate}}},
  \bibnamefont{and} \bibinfo{author}{\bibfnamefont{D.}~\bibnamefont{Sudarsky}},
  \bibinfo{journal}{Phys. Lett.} \textbf{\bibinfo{volume}{B743}},
  \bibinfo{pages}{484} (\bibinfo{year}{2015}), \eprint{1410.4212}.

\bibitem[{\citenamefont{Le\'on and Sudarsky}(2015)}]{Leon15}
\bibinfo{author}{\bibfnamefont{G.}~\bibnamefont{Le\'on}} \bibnamefont{and}
  \bibinfo{author}{\bibfnamefont{D.}~\bibnamefont{Sudarsky}},
  \bibinfo{journal}{JCAP} \textbf{\bibinfo{volume}{06}}, \bibinfo{pages}{020}
  (\bibinfo{year}{2015}), \eprint{1503.01417}.

\bibitem[{\citenamefont{Markkanen et~al.}(2015)\citenamefont{Markkanen,
  Rasanen, and Wahlman}}]{Syksy15}
\bibinfo{author}{\bibfnamefont{T.}~\bibnamefont{Markkanen}},
  \bibinfo{author}{\bibfnamefont{S.}~\bibnamefont{Rasanen}}, \bibnamefont{and}
  \bibinfo{author}{\bibfnamefont{P.}~\bibnamefont{Wahlman}},
  \bibinfo{journal}{Phys. Rev. D} \textbf{\bibinfo{volume}{91}},
  \bibinfo{pages}{084064} (\bibinfo{year}{2015}), \eprint{1407.4691}.

\bibitem[{\citenamefont{Leon and Bengochea}(2016)}]{Leon16}
\bibinfo{author}{\bibfnamefont{G.}~\bibnamefont{Leon}} \bibnamefont{and}
  \bibinfo{author}{\bibfnamefont{G.~R.} \bibnamefont{Bengochea}},
  \bibinfo{journal}{Eur. Phys. J.} \textbf{\bibinfo{volume}{C76}},
  \bibinfo{pages}{29} (\bibinfo{year}{2016}), \eprint{1502.04907}.

\bibitem[{\citenamefont{Alexander et~al.}(2016)\citenamefont{Alexander, Jyoti,
  and Magueijo}}]{Stephon16}
\bibinfo{author}{\bibfnamefont{S.}~\bibnamefont{Alexander}},
  \bibinfo{author}{\bibfnamefont{D.}~\bibnamefont{Jyoti}}, \bibnamefont{and}
  \bibinfo{author}{\bibfnamefont{J.}~\bibnamefont{Magueijo}},
  \bibinfo{journal}{Phys. Rev. D} \textbf{\bibinfo{volume}{94}},
  \bibinfo{pages}{043502} (\bibinfo{year}{2016}), \eprint{1602.01216}.

\bibitem[{\citenamefont{{Le{\'o}n}}(2017)}]{Leon17}
\bibinfo{author}{\bibfnamefont{G.}~\bibnamefont{{Le{\'o}n}}},
  \bibinfo{journal}{European Physical Journal C} \textbf{\bibinfo{volume}{77}},
  \bibinfo{eid}{705} (\bibinfo{year}{2017}), \eprint{1705.03958}.

\bibitem[{\citenamefont{{Landau} et~al.}(2012)\citenamefont{{Landau},
  {Sc{\'o}ccola}, and {Sudarsky}}}]{Landau12}
\bibinfo{author}{\bibfnamefont{S.~J.} \bibnamefont{{Landau}}},
  \bibinfo{author}{\bibfnamefont{C.~G.} \bibnamefont{{Sc{\'o}ccola}}},
  \bibnamefont{and}
  \bibinfo{author}{\bibfnamefont{D.}~\bibnamefont{{Sudarsky}}},
  \bibinfo{journal}{Physical Review D} \textbf{\bibinfo{volume}{85}},
  \bibinfo{eid}{123001} (\bibinfo{year}{2012}), \eprint{1112.1830}.

\bibitem[{\citenamefont{Benetti et~al.}(2016)\citenamefont{Benetti, Landau, and
  Alcaniz}}]{Benetti16}
\bibinfo{author}{\bibfnamefont{M.}~\bibnamefont{Benetti}},
  \bibinfo{author}{\bibfnamefont{S.~J.} \bibnamefont{Landau}},
  \bibnamefont{and} \bibinfo{author}{\bibfnamefont{J.~S.}
  \bibnamefont{Alcaniz}}, \bibinfo{journal}{JCAP}
  \textbf{\bibinfo{volume}{12}}, \bibinfo{pages}{035} (\bibinfo{year}{2016}),
  \eprint{1610.03091}.

\bibitem[{\citenamefont{Bengochea and Le\'on}(2017)}]{Bengo17}
\bibinfo{author}{\bibfnamefont{G.~R.} \bibnamefont{Bengochea}}
  \bibnamefont{and} \bibinfo{author}{\bibfnamefont{G.}~\bibnamefont{Le\'on}},
  \bibinfo{journal}{Phys. Lett. B} \textbf{\bibinfo{volume}{774}},
  \bibinfo{pages}{338} (\bibinfo{year}{2017}), \eprint{1708.07527}.

\bibitem[{\citenamefont{{Drossel} and {Ellis}}(2018)}]{Ellis18}
\bibinfo{author}{\bibfnamefont{B.}~\bibnamefont{{Drossel}}} \bibnamefont{and}
  \bibinfo{author}{\bibfnamefont{G.}~\bibnamefont{{Ellis}}},
  \bibinfo{journal}{New Journal of Physics} \textbf{\bibinfo{volume}{20}},
  \bibinfo{eid}{113025} (\bibinfo{year}{2018}), \eprint{1807.08171}.

\bibitem[{\citenamefont{Ca\~nate et~al.}(2018)\citenamefont{Ca\~nate, Ramirez,
  and Sudarsky}}]{Pedro18}
\bibinfo{author}{\bibfnamefont{P.}~\bibnamefont{Ca\~nate}},
  \bibinfo{author}{\bibfnamefont{E.}~\bibnamefont{Ramirez}}, \bibnamefont{and}
  \bibinfo{author}{\bibfnamefont{D.}~\bibnamefont{Sudarsky}},
  \bibinfo{journal}{JCAP} \textbf{\bibinfo{volume}{08}}, \bibinfo{pages}{043}
  (\bibinfo{year}{2018}), \eprint{1802.02238}.

\bibitem[{\citenamefont{Ju\'arez-Aubry
  et~al.}(2018)\citenamefont{Ju\'arez-Aubry, Kay, and Sudarsky}}]{Benito18}
\bibinfo{author}{\bibfnamefont{B.~A.} \bibnamefont{Ju\'arez-Aubry}},
  \bibinfo{author}{\bibfnamefont{B.~S.} \bibnamefont{Kay}}, \bibnamefont{and}
  \bibinfo{author}{\bibfnamefont{D.}~\bibnamefont{Sudarsky}},
  \bibinfo{journal}{Phys. Rev. D} \textbf{\bibinfo{volume}{97}},
  \bibinfo{pages}{025010} (\bibinfo{year}{2018}), \eprint{1708.09371}.

\bibitem[{\citenamefont{{Piccirilli} et~al.}(2019)\citenamefont{{Piccirilli},
  {Le{\'o}n}, {Landau}, {Benetti}, and {Sudarsky}}}]{Picci19}
\bibinfo{author}{\bibfnamefont{M.~P.} \bibnamefont{{Piccirilli}}},
  \bibinfo{author}{\bibfnamefont{G.}~\bibnamefont{{Le{\'o}n}}},
  \bibinfo{author}{\bibfnamefont{S.~J.} \bibnamefont{{Landau}}},
  \bibinfo{author}{\bibfnamefont{M.}~\bibnamefont{{Benetti}}},
  \bibnamefont{and}
  \bibinfo{author}{\bibfnamefont{D.}~\bibnamefont{{Sudarsky}}},
  \bibinfo{journal}{International Journal of Modern Physics D}
  \textbf{\bibinfo{volume}{28}}, \bibinfo{pages}{1950041}
  (\bibinfo{year}{2019}), \eprint{1709.06237}.

\bibitem[{\citenamefont{Le\'on et~al.}(2015)\citenamefont{Le\'on, Kraiselburd,
  and Landau}}]{Lucila15}
\bibinfo{author}{\bibfnamefont{G.}~\bibnamefont{Le\'on}},
  \bibinfo{author}{\bibfnamefont{L.}~\bibnamefont{Kraiselburd}},
  \bibnamefont{and} \bibinfo{author}{\bibfnamefont{S.~J.}
  \bibnamefont{Landau}}, \bibinfo{journal}{Phys. Rev. D}
  \textbf{\bibinfo{volume}{92}}, \bibinfo{pages}{083516}
  (\bibinfo{year}{2015}), \eprint{1509.08399}.

\bibitem[{\citenamefont{{Mariani} et~al.}(2016)\citenamefont{{Mariani},
  {Bengochea}, and {Le{\'o}n}}}]{Mariani16}
\bibinfo{author}{\bibfnamefont{M.}~\bibnamefont{{Mariani}}},
  \bibinfo{author}{\bibfnamefont{G.~R.} \bibnamefont{{Bengochea}}},
  \bibnamefont{and}
  \bibinfo{author}{\bibfnamefont{G.}~\bibnamefont{{Le{\'o}n}}},
  \bibinfo{journal}{Physics Letters B} \textbf{\bibinfo{volume}{752}},
  \bibinfo{pages}{344} (\bibinfo{year}{2016}), \eprint{1412.6471}.

\bibitem[{\citenamefont{Le\'on et~al.}(2017)\citenamefont{Le\'on, Majhi, Okon,
  and Sudarsky}}]{Maj17}
\bibinfo{author}{\bibfnamefont{G.}~\bibnamefont{Le\'on}},
  \bibinfo{author}{\bibfnamefont{A.}~\bibnamefont{Majhi}},
  \bibinfo{author}{\bibfnamefont{E.}~\bibnamefont{Okon}}, \bibnamefont{and}
  \bibinfo{author}{\bibfnamefont{D.}~\bibnamefont{Sudarsky}},
  \bibinfo{journal}{Phys. Rev. D} \textbf{\bibinfo{volume}{96}},
  \bibinfo{pages}{101301} (\bibinfo{year}{2017}), \eprint{1607.03523}.

\bibitem[{\citenamefont{Le\'on et~al.}(2018)\citenamefont{Le\'on, Majhi, Okon,
  and Sudarsky}}]{ModosB}
\bibinfo{author}{\bibfnamefont{G.}~\bibnamefont{Le\'on}},
  \bibinfo{author}{\bibfnamefont{A.}~\bibnamefont{Majhi}},
  \bibinfo{author}{\bibfnamefont{E.}~\bibnamefont{Okon}}, \bibnamefont{and}
  \bibinfo{author}{\bibfnamefont{D.}~\bibnamefont{Sudarsky}},
  \bibinfo{journal}{Phys. Rev. D} \textbf{\bibinfo{volume}{98}},
  \bibinfo{pages}{023512} (\bibinfo{year}{2018}), \eprint{1712.02435}.

\bibitem[{\citenamefont{{Le{\'o}n} et~al.}(2016)\citenamefont{{Le{\'o}n},
  {Bengochea}, and {Landau}}}]{Bouncing16}
\bibinfo{author}{\bibfnamefont{G.}~\bibnamefont{{Le{\'o}n}}},
  \bibinfo{author}{\bibfnamefont{G.~R.} \bibnamefont{{Bengochea}}},
  \bibnamefont{and} \bibinfo{author}{\bibfnamefont{S.~J.}
  \bibnamefont{{Landau}}}, \bibinfo{journal}{European Physical Journal C}
  \textbf{\bibinfo{volume}{76}}, \bibinfo{eid}{407} (\bibinfo{year}{2016}),
  \eprint{1605.03632}.

\bibitem[{\citenamefont{Josset et~al.}(2017)\citenamefont{Josset, Perez, and
  Sudarsky}}]{Josset17}
\bibinfo{author}{\bibfnamefont{T.}~\bibnamefont{Josset}},
  \bibinfo{author}{\bibfnamefont{A.}~\bibnamefont{Perez}}, \bibnamefont{and}
  \bibinfo{author}{\bibfnamefont{D.}~\bibnamefont{Sudarsky}},
  \bibinfo{journal}{Phys. Rev. Lett.} \textbf{\bibinfo{volume}{118}},
  \bibinfo{pages}{021102} (\bibinfo{year}{2017}), \eprint{1604.04183}.

\bibitem[{\citenamefont{Leon and Piccirilli}(2020)}]{Leon2020}
\bibinfo{author}{\bibfnamefont{G.}~\bibnamefont{Leon}} \bibnamefont{and}
  \bibinfo{author}{\bibfnamefont{M.~P.} \bibnamefont{Piccirilli}},
  \bibinfo{journal}{Phys. Rev. D} \textbf{\bibinfo{volume}{102}},
  \bibinfo{pages}{043515} (\bibinfo{year}{2020}), \eprint{2006.03092}.

\bibitem[{\citenamefont{Bengochea et~al.}(2021)\citenamefont{Bengochea,
  Piccirilli, and Le\'on}}]{BengoEmer21}
\bibinfo{author}{\bibfnamefont{G.~R.} \bibnamefont{Bengochea}},
  \bibinfo{author}{\bibfnamefont{M.~P.} \bibnamefont{Piccirilli}},
  \bibnamefont{and} \bibinfo{author}{\bibfnamefont{G.}~\bibnamefont{Le\'on}},
  \bibinfo{journal}{Eur. Phys. J. C} \textbf{\bibinfo{volume}{81}},
  \bibinfo{pages}{1049} (\bibinfo{year}{2021}), \eprint{2108.01472}.

\bibitem[{\citenamefont{Martin and Vennin}(2020)}]{MartinShadow}
\bibinfo{author}{\bibfnamefont{J.}~\bibnamefont{Martin}} \bibnamefont{and}
  \bibinfo{author}{\bibfnamefont{V.}~\bibnamefont{Vennin}},
  \bibinfo{journal}{Phys. Rev. Lett.} \textbf{\bibinfo{volume}{124}},
  \bibinfo{pages}{080402} (\bibinfo{year}{2020}), \eprint{1906.04405}.

\bibitem[{\citenamefont{Bengochea
  et~al.}(2020{\natexlab{a}})\citenamefont{Bengochea, Leon, Pearle, and
  Sudarsky}}]{Bengo20Letter}
\bibinfo{author}{\bibfnamefont{G.~R.} \bibnamefont{Bengochea}},
  \bibinfo{author}{\bibfnamefont{G.}~\bibnamefont{Leon}},
  \bibinfo{author}{\bibfnamefont{P.}~\bibnamefont{Pearle}}, \bibnamefont{and}
  \bibinfo{author}{\bibfnamefont{D.}~\bibnamefont{Sudarsky}},
  \bibinfo{journal}{arXiv e-prints} \bibinfo{eid}{arXiv:2006.05313}
  (\bibinfo{year}{2020}{\natexlab{a}}), \eprint{2006.05313}.

\bibitem[{\citenamefont{Bengochea
  et~al.}(2020{\natexlab{b}})\citenamefont{Bengochea, Le\'on, Pearle, and
  Sudarsky}}]{Bengo20Long}
\bibinfo{author}{\bibfnamefont{G.~R.} \bibnamefont{Bengochea}},
  \bibinfo{author}{\bibfnamefont{G.}~\bibnamefont{Le\'on}},
  \bibinfo{author}{\bibfnamefont{P.}~\bibnamefont{Pearle}}, \bibnamefont{and}
  \bibinfo{author}{\bibfnamefont{D.}~\bibnamefont{Sudarsky}},
  \bibinfo{journal}{Eur. Phys. J. C} \textbf{\bibinfo{volume}{80}},
  \bibinfo{pages}{1021} (\bibinfo{year}{2020}{\natexlab{b}}),
  \eprint{2008.05285}.

\bibitem[{\citenamefont{Martin and Vennin}(2021{\natexlab{a}})}]{Martin20R}
\bibinfo{author}{\bibfnamefont{J.}~\bibnamefont{Martin}} \bibnamefont{and}
  \bibinfo{author}{\bibfnamefont{V.}~\bibnamefont{Vennin}},
  \bibinfo{journal}{Eur. Phys. J. C} \textbf{\bibinfo{volume}{81}},
  \bibinfo{pages}{64} (\bibinfo{year}{2021}{\natexlab{a}}),
  \eprint{2010.04067}.

\bibitem[{\citenamefont{Gundhi et~al.}(2021)\citenamefont{Gundhi, Gaona-Reyes,
  Carlesso, and Bassi}}]{Bassi21}
\bibinfo{author}{\bibfnamefont{A.}~\bibnamefont{Gundhi}},
  \bibinfo{author}{\bibfnamefont{J.~L.} \bibnamefont{Gaona-Reyes}},
  \bibinfo{author}{\bibfnamefont{M.}~\bibnamefont{Carlesso}}, \bibnamefont{and}
  \bibinfo{author}{\bibfnamefont{A.}~\bibnamefont{Bassi}},
  \bibinfo{journal}{Phys. Rev. Lett.} \textbf{\bibinfo{volume}{127}},
  \bibinfo{pages}{091302} (\bibinfo{year}{2021}), \eprint{2102.07688}.

\bibitem[{\citenamefont{Martin and Vennin}(2021{\natexlab{b}})}]{Martin21}
\bibinfo{author}{\bibfnamefont{J.}~\bibnamefont{Martin}} \bibnamefont{and}
  \bibinfo{author}{\bibfnamefont{V.}~\bibnamefont{Vennin}},
  \bibinfo{journal}{Eur. Phys. J. C} \textbf{\bibinfo{volume}{81}},
  \bibinfo{pages}{516} (\bibinfo{year}{2021}{\natexlab{b}}),
  \eprint{2103.01697}.

\bibitem[{\citenamefont{Okon and Sudarsky}(2014)}]{Beneficios}
\bibinfo{author}{\bibfnamefont{E.}~\bibnamefont{Okon}} \bibnamefont{and}
  \bibinfo{author}{\bibfnamefont{D.}~\bibnamefont{Sudarsky}},
  \bibinfo{journal}{Found. Phys.} \textbf{\bibinfo{volume}{44}},
  \bibinfo{pages}{114} (\bibinfo{year}{2014}), \eprint{1309.1730}.

\bibitem[{\citenamefont{Modak et~al.}(2015{\natexlab{a}})\citenamefont{Modak,
  Ort\'iz, Pe\~na, and Sudarsky}}]{Modak14}
\bibinfo{author}{\bibfnamefont{S.~K.} \bibnamefont{Modak}},
  \bibinfo{author}{\bibfnamefont{L.}~\bibnamefont{Ort\'iz}},
  \bibinfo{author}{\bibfnamefont{I.}~\bibnamefont{Pe\~na}}, \bibnamefont{and}
  \bibinfo{author}{\bibfnamefont{D.}~\bibnamefont{Sudarsky}},
  \bibinfo{journal}{General Relativity and Gravitation}
  \textbf{\bibinfo{volume}{47}}, \bibinfo{eid}{120}
  (\bibinfo{year}{2015}{\natexlab{a}}), \eprint{1406.4898}.

\bibitem[{\citenamefont{Modak et~al.}(2015{\natexlab{b}})\citenamefont{Modak,
  Ort\'iz, Pe\~na, and Sudarsky}}]{Modak15}
\bibinfo{author}{\bibfnamefont{S.~K.} \bibnamefont{Modak}},
  \bibinfo{author}{\bibfnamefont{L.}~\bibnamefont{Ort\'iz}},
  \bibinfo{author}{\bibfnamefont{I.}~\bibnamefont{Pe\~na}}, \bibnamefont{and}
  \bibinfo{author}{\bibfnamefont{D.}~\bibnamefont{Sudarsky}},
  \bibinfo{journal}{Phys. Rev. D} \textbf{\bibinfo{volume}{91}},
  \bibinfo{eid}{124009} (\bibinfo{year}{2015}{\natexlab{b}}),
  \eprint{1408.3062}.

\bibitem[{\citenamefont{Bedingham et~al.}(2016)\citenamefont{Bedingham, Modak,
  and Sudarsky}}]{Modak16}
\bibinfo{author}{\bibfnamefont{D.}~\bibnamefont{Bedingham}},
  \bibinfo{author}{\bibfnamefont{S.~K.} \bibnamefont{Modak}}, \bibnamefont{and}
  \bibinfo{author}{\bibfnamefont{D.}~\bibnamefont{Sudarsky}},
  \bibinfo{journal}{Phys. Rev. D} \textbf{\bibinfo{volume}{94}},
  \bibinfo{eid}{045009} (\bibinfo{year}{2016}), \eprint{1604.06537}.

\bibitem[{\citenamefont{Corral et~al.}(2020)\citenamefont{Corral, Cruz, and
  Gonz\'alez}}]{Corral20}
\bibinfo{author}{\bibfnamefont{C.}~\bibnamefont{Corral}},
  \bibinfo{author}{\bibfnamefont{N.}~\bibnamefont{Cruz}}, \bibnamefont{and}
  \bibinfo{author}{\bibfnamefont{E.}~\bibnamefont{Gonz\'alez}},
  \bibinfo{journal}{Phys. Rev. D} \textbf{\bibinfo{volume}{102}},
  \bibinfo{pages}{023508} (\bibinfo{year}{2020}), \eprint{2005.06052}.

\bibitem[{\citenamefont{Linares Cede\~no and Nucamendi}(2021)}]{Nucamendi20}
\bibinfo{author}{\bibfnamefont{F.~X.} \bibnamefont{Linares Cede\~no}}
  \bibnamefont{and}
  \bibinfo{author}{\bibfnamefont{U.}~\bibnamefont{Nucamendi}},
  \bibinfo{journal}{Phys. Dark Univ.} \textbf{\bibinfo{volume}{32}},
  \bibinfo{pages}{100807} (\bibinfo{year}{2021}), \eprint{2009.10268}.

\bibitem[{\citenamefont{Eppley and Hannah}(1977)}]{Eppley77}
\bibinfo{author}{\bibfnamefont{K.}~\bibnamefont{Eppley}} \bibnamefont{and}
  \bibinfo{author}{\bibfnamefont{E.}~\bibnamefont{Hannah}},
  \bibinfo{journal}{Foundations of Physics} \textbf{\bibinfo{volume}{7}},
  \bibinfo{pages}{51} (\bibinfo{year}{1977}).

\bibitem[{\citenamefont{Page and Geilker}(1981)}]{Page81}
\bibinfo{author}{\bibfnamefont{D.~N.} \bibnamefont{Page}} \bibnamefont{and}
  \bibinfo{author}{\bibfnamefont{C.~D.} \bibnamefont{Geilker}},
  \bibinfo{journal}{Phys. Rev. Lett.} \textbf{\bibinfo{volume}{47}},
  \bibinfo{pages}{979} (\bibinfo{year}{1981}).

\bibitem[{\citenamefont{{Mattingly}}(2005)}]{Mattingly05}
\bibinfo{author}{\bibfnamefont{J.}~\bibnamefont{{Mattingly}}},
  \emph{\bibinfo{title}{{Is Quantum Gravity Necessary?}}}
  (\bibinfo{year}{2005}), vol.~\bibinfo{volume}{11}, pp.
  \bibinfo{pages}{327--338}.

\bibitem[{\citenamefont{Mattingly}(2006)}]{Mattingly06}
\bibinfo{author}{\bibfnamefont{J.}~\bibnamefont{Mattingly}},
  \bibinfo{journal}{Phys. Rev.} \textbf{\bibinfo{volume}{D73}},
  \bibinfo{pages}{064025} (\bibinfo{year}{2006}), \eprint{gr-qc/0601127}.

\bibitem[{\citenamefont{Kent}(2018)}]{Kent18}
\bibinfo{author}{\bibfnamefont{A.}~\bibnamefont{Kent}},
  \bibinfo{journal}{Class. Quant. Grav.} \textbf{\bibinfo{volume}{35}},
  \bibinfo{pages}{245008} (\bibinfo{year}{2018}), \eprint{1807.08708}.

\bibitem[{\citenamefont{Tilloy and Di\'{o}si}(2016)}]{Tilloy16}
\bibinfo{author}{\bibfnamefont{A.}~\bibnamefont{Tilloy}} \bibnamefont{and}
  \bibinfo{author}{\bibfnamefont{L.}~\bibnamefont{Di\'{o}si}},
  \bibinfo{journal}{Phys. Rev.} \textbf{\bibinfo{volume}{D93}},
  \bibinfo{pages}{024026} (\bibinfo{year}{2016}), \eprint{1509.08705}.

\bibitem[{\citenamefont{Carlip}(2008)}]{Carlip08}
\bibinfo{author}{\bibfnamefont{S.}~\bibnamefont{Carlip}},
  \bibinfo{journal}{Class. Quant. Grav.} \textbf{\bibinfo{volume}{25}},
  \bibinfo{pages}{154010} (\bibinfo{year}{2008}), \eprint{0803.3456}.

\bibitem[{\citenamefont{Albers et~al.}(2008)\citenamefont{Albers, Kiefer, and
  Reginatto}}]{Albers08}
\bibinfo{author}{\bibfnamefont{M.}~\bibnamefont{Albers}},
  \bibinfo{author}{\bibfnamefont{C.}~\bibnamefont{Kiefer}}, \bibnamefont{and}
  \bibinfo{author}{\bibfnamefont{M.}~\bibnamefont{Reginatto}},
  \bibinfo{journal}{Phys. Rev.} \textbf{\bibinfo{volume}{D78}},
  \bibinfo{pages}{064051} (\bibinfo{year}{2008}), \eprint{0802.1978}.

\bibitem[{\citenamefont{Ford}(2005)}]{Ford05Review}
\bibinfo{author}{\bibfnamefont{L.~H.} \bibnamefont{Ford}}
  (\bibinfo{year}{2005}), \eprint{gr-qc/0504096}.

\bibitem[{\citenamefont{{Ford}}(2005)}]{Ford05ReviewBook}
\bibinfo{author}{\bibfnamefont{L.~H.} \bibnamefont{{Ford}}},
  \emph{\bibinfo{title}{{Spacetime in Semiclassical Gravity}}}
  (\bibinfo{year}{2005}), pp. \bibinfo{pages}{293--310}.

\bibitem[{\citenamefont{Ju\'arez-Aubry
  et~al.}(2020)\citenamefont{Ju\'arez-Aubry, Miramontes, and
  Sudarsky}}]{Benito20}
\bibinfo{author}{\bibfnamefont{B.~A.} \bibnamefont{Ju\'arez-Aubry}},
  \bibinfo{author}{\bibfnamefont{T.}~\bibnamefont{Miramontes}},
  \bibnamefont{and} \bibinfo{author}{\bibfnamefont{D.}~\bibnamefont{Sudarsky}},
  \bibinfo{journal}{J. Math. Phys.} \textbf{\bibinfo{volume}{61}},
  \bibinfo{pages}{032301} (\bibinfo{year}{2020}), \eprint{1907.09960}.

\bibitem[{\citenamefont{Pearle and Squires}(1994)}]{Pearle1994}
\bibinfo{author}{\bibfnamefont{P.~M.} \bibnamefont{Pearle}} \bibnamefont{and}
  \bibinfo{author}{\bibfnamefont{E.}~\bibnamefont{Squires}},
  \bibinfo{journal}{Phys. Rev. Lett.} \textbf{\bibinfo{volume}{73}},
  \bibinfo{pages}{1} (\bibinfo{year}{1994}).

\bibitem[{\citenamefont{{Di{\'o}si}}(1984)}]{Diosi84}
\bibinfo{author}{\bibfnamefont{L.}~\bibnamefont{{Di{\'o}si}}},
  \bibinfo{journal}{Physics Letters A} \textbf{\bibinfo{volume}{105}},
  \bibinfo{pages}{199} (\bibinfo{year}{1984}), \eprint{1412.0201}.

\bibitem[{\citenamefont{Pearle and Squires}(1996)}]{Pearle1995}
\bibinfo{author}{\bibfnamefont{P.~M.} \bibnamefont{Pearle}} \bibnamefont{and}
  \bibinfo{author}{\bibfnamefont{E.}~\bibnamefont{Squires}},
  \bibinfo{journal}{Found. Phys.} \textbf{\bibinfo{volume}{26}},
  \bibinfo{pages}{291} (\bibinfo{year}{1996}), \eprint{quant-ph/9503019}.

\bibitem[{\citenamefont{Okon and Sudarsky}(2016{\natexlab{b}})}]{Okon16b}
\bibinfo{author}{\bibfnamefont{E.}~\bibnamefont{Okon}} \bibnamefont{and}
  \bibinfo{author}{\bibfnamefont{D.}~\bibnamefont{Sudarsky}},
  \bibinfo{journal}{Class. Quant. Grav.} \textbf{\bibinfo{volume}{33}},
  \bibinfo{pages}{225015} (\bibinfo{year}{2016}{\natexlab{b}}),
  \eprint{1602.07006}.

\bibitem[{\citenamefont{{Tumulka}}(2006)}]{Tumulka06}
\bibinfo{author}{\bibfnamefont{R.}~\bibnamefont{{Tumulka}}},
  \bibinfo{journal}{Journal of Statistical Physics}
  \textbf{\bibinfo{volume}{125}}, \bibinfo{pages}{821} (\bibinfo{year}{2006}),
  \eprint{quant-ph/0406094}.

\bibitem[{\citenamefont{Bedingham}(2011)}]{Bedingham11}
\bibinfo{author}{\bibfnamefont{D.~J.} \bibnamefont{Bedingham}},
  \bibinfo{journal}{Found. Phys.} \textbf{\bibinfo{volume}{41}},
  \bibinfo{pages}{686} (\bibinfo{year}{2011}), \eprint{1003.2774}.

\bibitem[{\citenamefont{Pearle}(2015)}]{Pearle2015}
\bibinfo{author}{\bibfnamefont{P.}~\bibnamefont{Pearle}},
  \bibinfo{journal}{Phys. Rev. D} \textbf{\bibinfo{volume}{91}},
  \bibinfo{pages}{105012} (\bibinfo{year}{2015}), \eprint{1412.6723}.

\bibitem[{\citenamefont{{Pearle}}(2012)}]{PearleMisc}
\bibinfo{author}{\bibfnamefont{P.}~\bibnamefont{{Pearle}}},
  \bibinfo{journal}{arXiv e-prints} \bibinfo{eid}{arXiv:1209.5082}
  (\bibinfo{year}{2012}), \eprint{1209.5082}.

\bibitem[{\citenamefont{Gasbarri et~al.}(2021)\citenamefont{Gasbarri,
  Belenchia, Carlesso, Donadi, Bassi, Kaltenbaek, Paternostro, and
  Ulbricht}}]{Bassiexp21}
\bibinfo{author}{\bibfnamefont{G.}~\bibnamefont{Gasbarri}},
  \bibinfo{author}{\bibfnamefont{A.}~\bibnamefont{Belenchia}},
  \bibinfo{author}{\bibfnamefont{M.}~\bibnamefont{Carlesso}},
  \bibinfo{author}{\bibfnamefont{S.}~\bibnamefont{Donadi}},
  \bibinfo{author}{\bibfnamefont{A.}~\bibnamefont{Bassi}},
  \bibinfo{author}{\bibfnamefont{R.}~\bibnamefont{Kaltenbaek}},
  \bibinfo{author}{\bibfnamefont{M.}~\bibnamefont{Paternostro}},
  \bibnamefont{and} \bibinfo{author}{\bibfnamefont{H.}~\bibnamefont{Ulbricht}},
  \bibinfo{journal}{Commun. Phys.} \textbf{\bibinfo{volume}{4}},
  \bibinfo{pages}{155} (\bibinfo{year}{2021}), \eprint{2106.05349}.

\bibitem[{\citenamefont{Lochan et~al.}(2012)\citenamefont{Lochan, Das, and
  Bassi}}]{BassiCMB}
\bibinfo{author}{\bibfnamefont{K.}~\bibnamefont{Lochan}},
  \bibinfo{author}{\bibfnamefont{S.}~\bibnamefont{Das}}, \bibnamefont{and}
  \bibinfo{author}{\bibfnamefont{A.}~\bibnamefont{Bassi}},
  \bibinfo{journal}{Phys. Rev. D} \textbf{\bibinfo{volume}{86}},
  \bibinfo{pages}{065016} (\bibinfo{year}{2012}), \eprint{1206.4425}.

\bibitem[{\citenamefont{Birrell and Davies}(1984)}]{Birrell}
\bibinfo{author}{\bibfnamefont{N.~D.} \bibnamefont{Birrell}} \bibnamefont{and}
  \bibinfo{author}{\bibfnamefont{P.~C.~W.} \bibnamefont{Davies}},
  \emph{\bibinfo{title}{{Quantum Fields in Curved Space}}}, Cambridge
  Monographs on Mathematical Physics (\bibinfo{publisher}{Cambridge Univ.
  Press}, \bibinfo{address}{Cambridge, UK}, \bibinfo{year}{1984}).

\bibitem[{\citenamefont{Mukhanov et~al.}(1992)\citenamefont{Mukhanov, Feldman,
  and Brandenberger}}]{mukhanov92}
\bibinfo{author}{\bibfnamefont{V.~F.} \bibnamefont{Mukhanov}},
  \bibinfo{author}{\bibfnamefont{H.~A.} \bibnamefont{Feldman}},
  \bibnamefont{and} \bibinfo{author}{\bibfnamefont{R.~H.}
  \bibnamefont{Brandenberger}}, \bibinfo{journal}{Phys. Rept.}
  \textbf{\bibinfo{volume}{215}}, \bibinfo{pages}{203} (\bibinfo{year}{1992}).

\bibitem[{\citenamefont{Mukhanov}(2005)}]{mukhanov2005}
\bibinfo{author}{\bibfnamefont{V.}~\bibnamefont{Mukhanov}},
  \emph{\bibinfo{title}{Physical Foundations of Cosmology}}
  (\bibinfo{publisher}{New York: Cambridge University Press},
  \bibinfo{year}{2005}).

\end{thebibliography}
\bibliographystyle{apsrev}

\end{document}